\documentstyle[prd,twocolumn,aps,floats]{revtex} 
\begin{document}
\input epsf
\def\be{\begin{equation}}\def\ee{\end{equation}}
\def\bea{\begin{eqnarray}}\def\eea{\end{eqnarray}}

\def\goesas{\mathop{\sim}\limits} \def\LA{\Lambda} \def\etal{{\it et al.}}
\def\si{\sigma} \def\rh{\rho} \def\ph{\phi} \def\qn{q_0} \def\al{\alpha}
\def\Om{\Omega} \def\gm{\gamma} \def\kp{\kappa} \def\ff{f} \def\FF{{\cal F}}
\def\rarr{\rightarrow} \def\mfac{\sqrt{\frac23}\,m} \def\CC{{\cal C}}
\def\const{\hbox{const}} \def\Ho{{H_1}} \def\w#1{\;\hbox{#1}\;}
\def\Ao{{\rm A}} \def\Bo{{\rm B}} \def\Co{{\rm C}} \def\e{{\rm e}}
\def\lsim{\mathop{\hbox{${\lower3.8pt\hbox{$<$}}\atop{\raise0.2pt\hbox{$\sim$}}
$}}} \def\ns#1{_{\text{#1}}} \def\meff#1{m^{\text{eff}}_{#1}}
\def\lsim{\mathop{\hbox{${\lower3.8pt\hbox{$<$}}\atop{\raise0.2pt\hbox{$\sim$}}
$}}} \def\gsim{\mathop{\hbox{${\lower3.8pt\hbox{$>$}}\atop{\raise0.2pt\hbox{$
\sim$}}$}}} \def\GG{{\cal G}} \def\HH{{\cal H}} \def\MM{{\cal M}_0}
\def\Omp{\Om_{\ph0}} \def\bsi{{\bar\si}}
\wideabs{
\title{Properties of cosmologies with dynamical
pseudo Nambu--Goldstone bosons}
\author{S.C. Cindy Ng\cite{Ecng} and David L. Wiltshire\cite{Edlw}}
\address{Department of Physics and Mathematical Physics, University of
Adelaide,\break Adelaide, S.A. 5005, Australia.}
\date{11 April, 2000; ADP-00-12/M91, astro-ph/0004138; Phys. Rev. {\bf D 63},
023503 (2001).}
\maketitle

\begin{abstract}
We study observational constraints on cosmological models with a quintessence
field in the form of a dynamical pseudo Nambu--Goldstone boson. After
reviewing the properties of the solutions, from a dynamical systems phase
space analysis, we consider the constraints on parameter values imposed by
luminosity distances from the 60 type Ia supernovae published by Perlmutter
\etal, and also from gravitational lensing statistics of distant quasars.
In the case of the type Ia supernovae we explicitly allow for the possibility
of evolution of the peak luminosities of the supernovae sources, using
simple empirical models which have been recently discussed in the literature.
We find weak evidence to suggest that the models with supernovae evolution
fit the data better in the context of the quintessence models in question.
If source evolution is a reality then the greatest challenge facing these
models is the tension between current value of the expansion age, $H_0t_0$,
and the fraction of the critical energy density, $\Omp$, corresponding
to the scalar field. Nonetheless there are ranges of the free parameters
which fit all available cosmological data.
\end{abstract}
}
\narrowtext

\section{Introduction}

Scalar fields have played a central role in models of the very early universe
for the past 20 years. In the past few years attention has turned to models in
which a scalar field plays a dynamical role at late times, rather than simply
being frozen in as a static relic vacuum energy. Such models, which have been
dubbed ``quintessence'' models \cite{CDS}, could in principle provide a
dynamical solution to the cosmological constant problem -- namely the question
of why the magnitude of the vacuum energy at the present epoch is so much
smaller than one might na\"{\i}vely expect from particle physics models such as
various supergravity theories. A dynamical `solution' of the cosmological
constant problem would amount to a demonstration that a particular dynamical
evolution of the scalar quintessence field is a natural consequence of the
cosmological field equations without fine-tuning of parameters, given some
reasonable physical assumptions about the initial conditions.

The most notable recent observational evidence which has driven the theoretical
interest is the measurement of the apparent magnitude-redshift relationship
using type Ia supernovae (SNe Ia) \cite{PS}. These results have been
interpreted, in the context of a cosmological model containing pressureless
dust and a cosmological constant, $\LA$, as evidence that the universe is
undergoing accelerated expansion at the present epoch (see \cite{P98,R98} and
references therein). The validity of this
conclusion is currently open to some doubt, however. In particular, a recent
analysis by Riess \etal\ \cite{Rie1} indicates that the sample of type Ia
supernovae shows a possible evolution in rise times from moderate ($z\goesas
0.3$) to large ($z\goesas1$) redshifts. Although the statistical significance
of this result has been diminished -- from the $5.8\si$ level \cite{Rie1} to
the $1.5\si$ level \cite{AKN} -- upon a more rigorous treatment of the
uncertainties in the data \cite{AKN}, it remains true that while a systematic
evolution in the rise times of the supernovae is not conclusively ruled in,
neither is it conclusively ruled out.

Given that an evolution in the shape of the light curves of the supernovae
measured in their rest frame remains a real possibility, it would not be
surprising if the
peak luminosity -- which is the effective standard candle used -- were also to
evolve. Riess \etal\ \cite{Rie1} conclude that the type Ia supernovae data
could conceivably be explained entirely within the context of an open
Friedmann-Robertson-Walker universe together with a reasonable astrophysical
evolution model, e.g., a consequence of a time variation of the abundances of
relevant heavy elements in the environment of the white dwarf supernovae
progenitors. Detailed astrophysical modelling -- see, e.g., \cite{Arn} --
should hopefully eventually resolve the issue, although at this stage the
difference between our theoretical understanding and the observations
remains quite substantial \cite{Rie2}.

In many recent papers it has been commonly assumed that the dynamical scalar
field, $\ph$, should obey an effective equation of state $P_\ph\simeq w\rh_\ph$
with $-1<w<0$, at the present epoch, in order to obtain a cosmological
acceleration, i.e., a negative deceleration parameter $\qn$. Indeed, the
condition that $-1<w<0$ is often taken as a defining characteristic of
``quintessence'' \cite{CDS}. The broad picture in this cosmological scenario is
that the universe is currently in the early stage of an epoch of inflationary
expansion. The motivation for this is that one could then hope to have a model
cosmology in which observations such as the type Ia supernovae apparent
magnitude--redshift relation could be explained by a cosmological acceleration
in a similar fashion to models with a cosmological constant, but with the
possibility of explaining why the magnitude of the vacuum energy density and
the energy density in ordinary pressureless matter, $\rh_m$, are comparable at
the present epoch -- the so-called ``cosmic coincidence problem'' \cite{ccp}.

One attractive feature of homogeneous isotropic cosmological models with
dynamical scalar fields is that many of them possess ``cosmological scaling
solutions'' \cite{LS}, namely solutions which at late times have energy density
components which depend on the cosmic scale factor according to $\rh\propto a^
{-m_1}$ and $\rh_\ph\propto a^{-m_2}$ simultaneously, and which act as
attractors in the phase space. If $m_2<m_1$, which is the case, for example,
for simple power-law potentials with inverse powers \cite{PR,ZWS}
$V(\ph)\propto \ph^{-\al}$, or for certain power-law potentials with positive
powers \cite{LS} then the scalar field dominates at late times, producing a
quintessence-dominated cosmology with accelerated expansion at late times. If
$m_1=m_2$, which is the case for exponential potentials
\cite{expV1,expV2,WCL,FJ,VL,FW,CLW,MP}, then the scaling solutions are
``self-tuning'' \cite{FJ} -- i.e., the dynamics of the scalar field follows
that of the other dominant energy component, with a dependence $\rh_\ph\propto
a^{-4}$ in the radiation--dominated era and a dependence $\rh_\ph\propto
a^{-3}$ in the matter--dominated era.

Even if the ultimate late-time behaviour properties of the solutions is
not precisely ``self-tuning'' in the above sense, models such as those with
inverse power law potentials can still effectively act as ``tracking
solutions'', since for a wide range of initial conditions, the
solutions rapidly converge to a common, cosmic evolutionary track \cite{ZWS}.
Thus there are a number of ways in which one might hope to
solve the ``cosmic coincidence problem'', though in practice a degree
of tuning of the parameters has been necessary in all models studied to date.

In this paper, we consider a form of quintessence, an ultra-light pseudo
Nambu-Goldstone-boson (PNGB) \cite{HR} which is still relaxing to its vacuum
state. From the viewpoint of quantum field theory PNGB models are the
simplest way to have naturally ultra-low mass, spin-$0$ particles and hence
perhaps the most natural candidate for a presently-existing minimally-coupled
scalar field. The effective potential of a PNGB field $\ph$ can be taken to
be of the form \cite{FHSW}
\be\label{VPNGB}V(\ph)=M^4[\cos(\ph/\ff)+1]\ ,\ee
where the constant term is to ensure that the vacuum energy vanishes at the
minimum of the potential. This potential is characterized by two mass scales, a
purely spontaneous symmetry breaking scale $\ff$ and an explicit symmetry
breaking scale $M$.

The effective PNGB mass is $m_\ph\goesas M^2/\ff$. To obtain solutions
with $\Om_\ph\goesas1$, the energy scales are essentially fixed \cite{HR}
to values $M\goesas10^{-3}\w{eV}$, interestingly close
to the neutrino mass scale for the MSW solution to the solar neutrino problem,
and $\ff\goesas m_{PL}\simeq10^{19}\w{GeV}$, the Planck scale. Since these two
energy scales have values which are reasonable from the viewpoint of particle
physics, one might hope to explain the coincidence that the vacuum energy is
dynamically important at the present epoch.

The cosmology of PNGB models has already been extensively studied in the
literature \cite{VL,FW,MP,FHSW,CDF,FW2}. In particular, a number of constraints
have been placed on the parameters $M$ and $\ff$ by various sets of
observational data \cite{VL,FW,CDF,FW2}. Most recently, Frieman and Waga
\cite{FW2} have set bounds based on the SNe Ia data of Riess \etal \cite{R98}
(hereafter R98) on the one hand, and gravitational lensing surveys on the
other. Comparing these bounds is of interest, since the SnIa data have been
interpreted as favouring a cosmological constant, whereas gravitational
lensing data has been used to place upper bounds on $\Lambda$
\cite{K95,Fal,Hel}. They therefore provide complementary tests of the
parameter spaces of models with a non-trivial vacuum energy.

In this paper it is our intention to critically study these bounds. Firstly,
we will consider how the bounds are affected by the initial value of the scalar
field at the beginning of the matter--dominated epoch. Secondly, we wish to
investigate how such bounds might be affected
in the case of the PNGB model if the observed apparent faintness of type Ia
supernovae is at least partly due to an intrinsic evolution of the sources
over cosmological time scales, which in view of the results of \cite{Rie1}
would appear to be a very real possibility. The reason for focusing on the
PNGB models in such an investigation is suggested by the fact that whereas many
quintessence models have been singled out in the literature, perhaps somewhat
artificially, simply because they have the property of yielding an accelerated
expansion, many different possibilities arise in the PNGB case. Indeed, at
very late times, the apparent magnitude--redshift relation for PNGB models
ultimately coincides with that of the Einstein--de Sitter model, even
though the density of ordinary matter can be low in the PNGB cosmologies. The
requirement that the faintness of type Ia supernovae is entirely due to
their cosmological distances places rather strong restrictions on the values
of the parameters $M$ and $\ff$ \cite{FW2}, because it requires us to exist at
an epoch of the PNGB cosmologies which is still quite far removed from our
ultimate destiny. If these restrictions are relaxed because of evolutionary
effects, then it is quite plausible that other regions of the parameter
space of the PNGB models become viable alternatives. Since PNGB cosmologies
could therefore still solve the ``missing energy problem'', even if the
evidence for a cosmological acceleration proves to be ephemeral, we believe
it is important to investigate this possibility quantitatively.

We will begin the paper with a qualitative analysis of the solutions, to
provide some general insights which will help to guide our quantitative
discussion. Although these properties are no doubt already known, to the
best of our knowledge an analysis of the phase space of the solutions
has never been presented in the literature. Having completed this analysis
in Sec.\ II we will go on to discuss a number of issues relating to numerical
integration in Sec.\ III, and relate the properties of the solutions
found numerically to the exact analysis of Sec.\ II. In Sec.\ IV we present
the main analysis of the constraints imposed on the $(M,\ff)$ parameter space,
allowing for the possibility of evolution of peak luminosities in the
type Ia supernova sources. Bounds from gravitational lensing statistics are
updated in Sec.\ V, and the implications of our results are discussed at
greater length in Sec. VI.

\section{Phase-Space Analysis}

We will begin by performing an analysis of the differential equations governing
the cosmological evolution in a manner similar to previous studies in
inflationary and quintessential models \cite{LS,expV2,WCL,CLW,MP}.

The classical action for gravity coupled to a scalar field $\ph$ has the form
\be\label{action}S=\int d^4x\sqrt{-g}\left[\left({R\over2\kp^2}-{1\over2}g^
{\mu\nu}\partial_\mu\ph\,\partial_\nu\ph-V(\ph)\right)+{\cal L}\right],\ee
where $\kp$ is the Planck constant, $R$ is the Ricci scalar,
$g\equiv\det{g_{\mu\nu}}$, and $L$ is the Lagrangian density of
non-relativistic matter and radiation. For simplicity, we assume $\ph$ is
minimally coupled to the curvature, and we work in units in which $\hbar=c=1$.

Consider a spatially-flat Friedmann-Robertson-Walker (FRW) universe containing
a fluid with barotropic equation of state $P_\gm=(\gm-1)\rh_\gm$, where $\gm$
is a constant, $0\leq\gm\leq2$, such as radiation ($\gm=4/3$) or dust
($\gm=1$). There is a self-interacting scalar field with the PNGB potential
energy density (\ref{VPNGB}) evolving in this universe. The total energy
density of this homogeneous scalar field is $\rh_\ph=\dot{\ph}^2/2+V(\ph)$. The
governing equations are given by
\bea\dot{H}&=&-{\kp^2\over2}\left(\rh_\gm+P_\gm+\dot{\ph}^2\right)\ ,\label
{Hdot}\\ \dot{\rh_\gm}&=&-3H(\rh_\gm+P_\gm )\ ,\label{rhodot}\\ \ddot{\ph}&=&-
3H\dot{\ph}-{dV\over d\ph}\ ,\label{phiddot}\eea
subject to the Friedmann constraint
\be\label{frecon}H^2={\kp^2\over3}\left(\rh_\gm+{1\over2}\dot{\ph}^2+V\right)
\ ,\ee
where $\kp^2\equiv8\pi G$, $H=\dot{a}/a$ is the Hubble parameter, and an
overdot denotes ordinary differentiation with respect to time $t$.

We may rewrite the Friedmann constraint as
\be\Om_\gm+\Om_\ph=1\ee
where
\bea\Om_\gm&=&{\kp^2\rh_\gm\over3H^2}\ ,\\ \Om_\ph&=&{\kp^2\over3H^2}\left({1
\over2}\dot{\ph}^2+V\right)\eea
are the ratios of the energy densities of the barotropic matter and the
quintessence field as fraction of the critical density respectively.

In contrast to the case of the cosmologies with an exponential potential
\cite{expV2,WCL,CLW} where the dynamics can be reduced to a 2-dimensional
autonomous phase plane, for the system (\ref{Hdot}) - (\ref{frecon}) the
simplest phase space appears to be 3-dimensional in the full four-dimensional
phase space.

There are two alternative choices of variables which are useful to describe the
dynamics, which we will discuss in turn.

\subsection{Field Variables}

The first choice is to simply use the Hubble parameter, H, the scalar field and
its first derivative as the elementary variables. These are of course simply
the variable $u$, $v$, $w$ of Frieman and Waga \cite{FW} up to an overall
scaling. By defining
\be\label{IJ}I\equiv{\ph\over\ff}={\kp\ph\over\FF}\ ;\qquad J\equiv\kp\dot{\ph
}\ ;\qquad\ee
we therefore obtain the system
\bea\dot{H}&=&-{3\gm\over2}H^2+{\gm m^2\over2}(\cos{I}+1)-{(2-\gm)\over4}J^2\
,\label{HdotIJ}\\ \dot{I}&=&{J\over\FF}\ ,\label{Idot}\\ \dot{J}&=&-3HJ+{m^2
\over\FF}\sin{I}\ ,\label{Jdot}\eea
where for notational simplicity we define $m^2=\kp^2M^4$, and $\FF=\kp\ff$.
so that $\FF$ is dimensionless, while $m$ has dimensions inverse time.
The constraint equation becomes
\be\label{FcHIJ}\kp^2\rh_\gm=3H^2-m^2(\cos{I}+1)-{1\over2}J^2\ .\ee

From Eq.~(\ref{rhodot}), it follows that $\dot{\rh_\gm}=0$ if $\rh_\gm=0$.
Therefore trajectories do not cross the 2-dimensional $\rh_\gm=0$ surface,
which is a hyperboloid in the variables $H$, $\cos(I/2)$, and $J$. Physical
trajectories with $\rh_\gm>0$ are forced to lie within the volume of the $H$,
$I$, $J$ phase space bounded by the $\rh_\gm=0$ surface.

The only critical points of the system (\ref{HdotIJ})-(\ref{Jdot}) at finite
values of $H$, $I$, $J$ occur at \begin{enumerate} \item $\Co_{1\pm}$ at
$H=\pm\Ho$, $I=0$ mod $2\pi$, $J=0$; and \item $\Co_2$ at $H=0$, $I=\pi$
mod $2\pi$, $J=0$, \end{enumerate} where
\be \Ho\equiv\mfac \label{Ho}\ee
Both of these points in fact lie on the $\rh_\gm=0$ surface. Furthermore, this
surface intersects the $H=0$ plane only at the isolated points $\Co_2$. The $H
>0$ and $H<0$ subspaces are thus physically distinct, and the $H<0$ subspace
simply correspond to the time-reversal of the $H>0$ subspace. Therefore we can
take $H>0$ without loss of generality.

The pattern of trajectories close to the $\rh_\gm=0$ surface can be ascertained
by continuity to the $\rh_\gm=0$ solutions, even though the latter are not
physical. The $\rh_\gm=0$ subspace is obtained, for example, by regarding
(\ref{FcHIJ}) as a quadratic equation for $H$, and using the solution to
eliminate $H$, thereby obtaining a 2-dimensional system for $I$ and $J$ given
by (\ref{Idot}) and (\ref{Jdot}).

\begin{figure}[htp] \centering\leavevmode\epsfysize=7cm \epsfbox{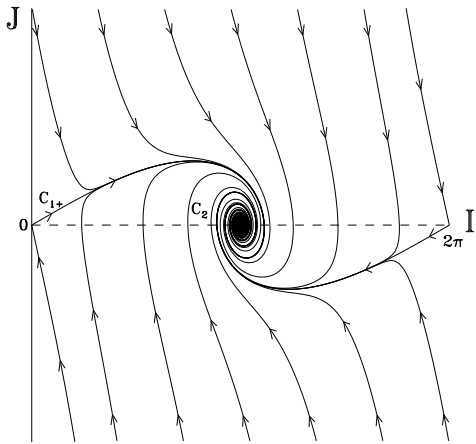}
\caption[rhgm0]{\label{rhgm0} The projection of the trajectories within the
$\rh_\gm=0$ subspace on the $I$ -- $J$ plane for values of $I\in[0,2\pi)$.
Within this subspace, $\Co_{1+}$ is a saddle point and $\Co_2$ is a stable
spiral.} \end{figure}

We plot the resulting $H>0$ pattern of trajectories in Fig.~\ref{rhgm0} for
values of $I\in[0,2\pi)$. Since the potential, $V(\ph)$, is periodic the same
pattern of trajectories repeats itself as we extend $I$ to $\pm\infty$, with
trajectories crossing from one ``cell'' to another at the cell boundaries.

The trajectories which occupy the lower half of Fig.~\ref{rhgm0} are obtained
from those in the upper half by the symmetry $I\rarr2\pi-I$, $J
\rarr-J$ of the differential equations (\ref{HdotIJ}) - (\ref{FcHIJ}).
Physically this simply corresponds to the scalar field rolling from the maximum
to the right of the minimum as opposed to the one to the left.

An analysis of small perturbations about the critical points $\Co_1$ and $\Co_2
$ yield eigenvalues \begin{enumerate} \item $\lambda=-\sqrt{6}m\gm,-{m\over2}
\left(\sqrt{6}\pm\sqrt{6+4/\FF^2}\right)$ at $\Co_{1+}$, \item$\lambda=0,\pm{im
\over\FF}$ at $\Co_2$. \end{enumerate}
Thus $\Co_{1+}$ attracts a two-dimensional
bunch of trajectories but is a saddle point with respect to trajectories lying
in the $\rh_\gm=0$ surface, as is evident from Fig.~\ref{rhgm0}. The
2-dimensional bunch of trajectories which approach $\Co_{1+}$ are found to
correspond to inflationary solution with $a\propto\exp(\sqrt{2/3}mt)$ as $t
\rarr\infty$ and $\ph\rarr$ const $=2n\pi\FF$, $n\in$ Z. The possible role of
scalar fields with PNGB potentials in driving an inflationary expansion of the
early universe has been discussed in \cite{ni}.

The point $\Co_2$ is a degenerate case, in particular with regard to
perturbations orthogonal to the $\rh_\gm=0$ surface, (i.e.\ into the surface
$\rh_\gm>0$ region), for which the eigenvalue is zero. It is a centre with
respect to the trajectories lying in the $\rh_\gm=0$ surface, and when
perturbations of higher order are considered it becomes a stable spiral point
in the $\rh_\gm=0$ surface as can be seen in Fig.~\ref{rhgm0}. Since there is a
degeneracy, however, an alternative choice of phase space variables is
desirable. We will defer a discussion of the late time behaviour of the
solution near $\Co_2$ to Sec.~\ref{Evars}.

The points $\Co_{1\pm}$ correspond to models with a scalar field sitting at the
maximum of the potential, whereas $\Co_2$ corresponds to the scalar field
sitting at the bottom of the potential well. The separatrices in Fig.\
\ref{rhgm0} which join $\Co_{1\pm}$ to $\Co_2$ correspond to the field
rolling from the maximum to the minimum. It would appear from Fig.~\ref{rhgm0}
trajectories which spiral into $\Co_2$ become arbitrarily close to the
separatrix at late times.

The separatrices which join the points $\Co_{1\pm}$ to points at
infinity correspond to solutions for which the scalar field reaches the
top of the potential hill as $t\rarr\pm\infty$, (e.g., the rightmost
trajectory in Fig.~\ref{rhgm0}). Finally, there are also straight line
separatrices parallel to the $H$-axis at each of the points $\Co_{1\pm}$,
extending from $H=\pm H_1$ to infinity,
which represent solutions with a static scalar field sitting on top of the
potential hill.

To examine the critical points at infinity it is convenient to transform to
spherical polar coordinates $r$, $\theta$, and $\ph$ by defining
\bea\label{pcH}H&=&r\cos\theta\ ,\\ \label{pcI}I&=&r\sin\theta\sin\ph\ ,\\
\label{pcJ}J&=&r\sin\theta\cos\ph,\eea
and to bring the sphere at infinity to a finite distance from the origin by the
transformation $r=\rh/(1-\rh )$, $0\leq\rh\leq1$ \cite{Bel}.

\begin{figure*}[htp] \centering \leavevmode\epsfysize=8cm \epsfbox{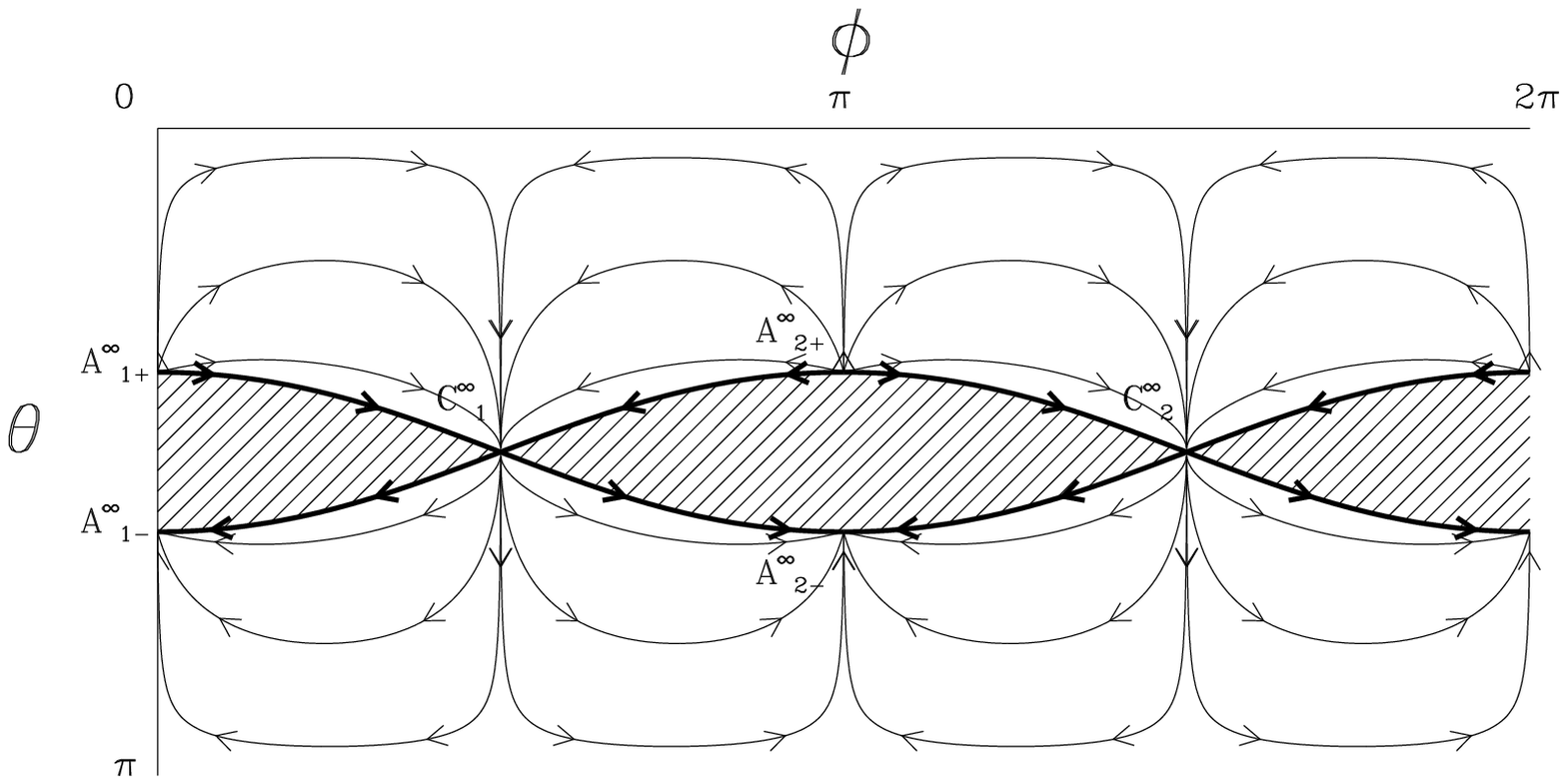}
\caption[rho1]{\label{rho1} The projection of trajectories within the sphere
at infinity on the $\ph$ -- $\theta$ plane. The unphysical region is shaded.}
\end{figure*}

Although the trajectories on the sphere at infinity do not represent physical
cosmologies, it is useful to plot them since the form of the trajectories which
lie just within the sphere will be similar. On the sphere $\rh=1$ we find
\bea{d\theta\over d\xi}&=&\sin\theta\cos^2\theta\left[3\sin^2\ph+{2-\gm\over
4}(\tan^2\theta\cos^2\ph-6)\right],\nonumber\\ \\
{d\ph\over d\xi}&=&{3\over2}\cos\theta\sin{2\ph},\eea
where $\xi$ is a new time coordinate defined by $d\xi=rdt$. The resulting
integral curves are plotted in Fig.~\ref{rho1}. By (\ref{FcHIJ}) the projection
of the physical region $\rh_\gm>0$ onto the sphere at infinity leads to the
condition
\be\cot^2\theta>{1\over6}\cos^2\ph.\ee
Values of $\theta$ and $\ph$ which violate this inequality lie in the shaded
region.

\begin{figure*}[htp] \centering \leavevmode\epsfysize=8cm
\epsfbox{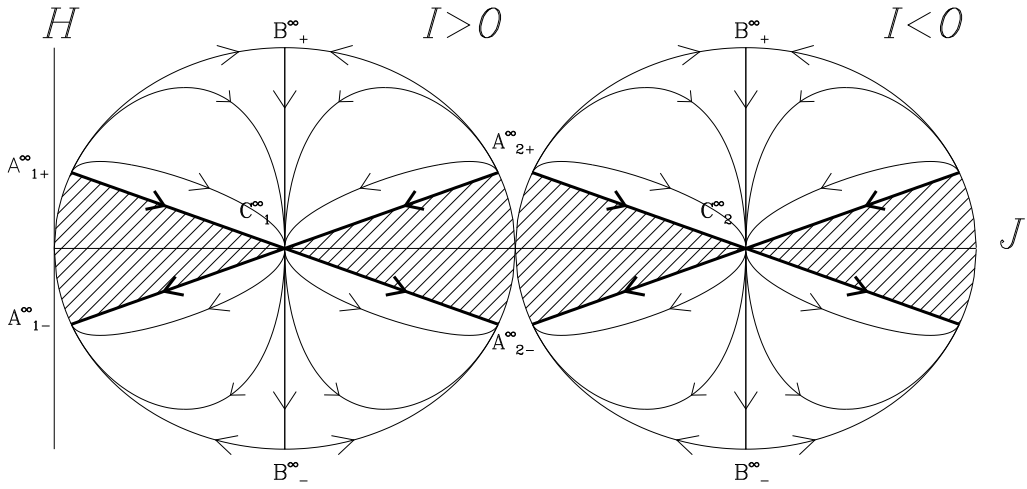} \caption[rho1hp]{\label{rho1hp} The projection of
trajectories within the $I>0$ and $I<0$ hemispheres at infinity on the $H$ --
$J$ plane. The unphysical region is shaded.} \end{figure*}

The critical points on the sphere at infinity are \begin{enumerate} \item
$\Ao_{1\pm}^\infty$, $\Ao_{2\pm}^\infty$: four points at

$(\theta,\ph)\in\{(\pm\tan^{-1}\sqrt{6},0),(\pm\tan^{-1}\sqrt{6},\pi)\}$ or

$H=\pm\infty$, $I/H=0$, and $J/H=\pm\sqrt{6}$.

\item $\Bo_{\pm}^\infty$: two points at

$H=\pm\infty$, $I/H=0$, and $J/H=0$.

Since the projection onto spherical polar coordinates (\ref{pcH}) - (\ref{pcJ})
is degenerate at the north and south poles $\theta=0,\pi$, these points are
excluded from the chart $(\theta,\ph)$ but can be included using an alternative
hemispherical projection (see Fig.~\ref{rho1hp}).

\item $\Co_{1,2}^\infty$: two points at

$(\theta,\ph)\in\{(\pi/2,\pi/2),(\pi/2,3\pi/2)\}$ or

$I=\pm\infty$, $H/I=0$, and $J/I=0$. \end{enumerate}

\begin{table}[hb] \centering \caption[eival]{\label{eival} The critical points
on the sphere at infinity and their eigenvalues.} \bigskip
\begin{tabular}{l|l} \noalign{\smallskip} Critical points & Eigenvalues (with
degeneracies) \\ \noalign{\smallskip} \hline \noalign{\smallskip}
$\Ao_{1\pm}^\infty$, $\Ao_{2\pm}^\infty$ & $\pm3$ (2), $\pm3(2-\gm)$ \\ \\
$\Bo_{\pm}^\infty$ & $\pm{3\gm\over2}$ (2), $\mp{3\over2}(2-\gm)$ \\ \\
$\Co_{1,2}^\infty$ & 0 (3) \\ \noalign{\smallskip} \end{tabular} \end{table}

An analysis of small perturbations shows that the points $\Ao_{1+,2+}^\infty$
are repellors in all directions of the phase space (cf., Table~\ref{eival}),
while $\Ao_{1-,2-}^\infty$ are attractors. This therefore represents the most
`typical'
early behaviour of solutions. For $H>0$, $\Ao_{1+,2+}^\infty$ correspond to the
limit $t\rarr0$. We find that $H\goesas{1\over3t}$ or $a\propto t^{1/3}$,
while $\kp\ph\goesas\pm\sqrt{2/3}\ln t$, for these solutions. The points $\Ao_
{1-,2-}^\infty$ with $H<0$ represent the time-reversed solutions.

The point $\Bo_+^\infty$ ($\Bo_-^\infty$) repels (attracts) a 2-dimensional
bunch of trajectories travelling to (from) finite values of $H$, $I$, $J$,
but is a saddle point with respect to directions on the sphere at infinity.
The points are found to correspond to $t\rarr0$ with $H\goesas{2\over3\gm t}$
or $a\propto t^{2/3\gm}$ while $\kp\ph\propto t^n$, $n>0$.

The points $\Co_{1,2}^\infty$ are the projection of the points $\Co_{1\pm}$ and
$\Co_2$ into the sphere at infinity. The degenerate eigenvalues simply reflect
the degeneracy of the projection.

$\Bo_+^\infty$ acts as a repellor for trajectories with $\dot{\ph}\simeq0$,
$\ph\simeq$ const as $t\rarr0$. As shown in Fig.~\ref{rho1hp}, trajectories
are driven towards $\Bo_+^\infty$ before they reach $\Co_{1}^\infty$. This
is consistent with the property that when $H$ is large ($3H\ge m_\ph$), the
field evolution is over-damped by the expansion, and the field is effectively
frozen to its initial value ($\dot{\ph}\rarr0$).

\subsection{Energy Density Variables} \label{Evars}

In view of the eigenvalue degeneracy encountered above, we can alternatively
choose to represent the system by the variables $H$, $x$, and $y$, where
\bea x&\equiv&{\kp\dot{\ph}\over\sqrt{6}H}={J\over\sqrt{6}H}\ ,\\ y&\equiv&
{\kp\sqrt{V}\over\sqrt{3}H}={\sqrt{2}m\cos(I/2)\over\sqrt{3}H}\ .\eea
These are the same variables used by Copeland, Liddle and Wands \cite{CLW} for
the model with an exponential potential. As above, we will consider $H>0$ only.
The field equations then take the form
\bea\label{Hdotxy}\dot{H}&=&-{3\over2}H^2\mu\ ,\\ \label{xdot}
\dot{x}&=&\pm{my \over\FF}\sqrt{1-{3y^2H^2\over2m^2}}+{3\over2}Hx(\mu-2)\ ,\\
\label{ydot}\dot{y}&=&\mp{mx\over\FF}\sqrt{1-{3y^2H^2\over2m^2}}+{3\over2}Hy
\mu\ ,\eea
where
\be\label{mu}\mu(x,y)\equiv\gm(1-y^2)+(2-\gm)x^2\ .\ee
We note that in these variables
\be x^2+y^2=\Om_\ph\label{Omphidef}\ee
which is why we have adopted the terminology ``energy density variables''. The
physical region of the phase space will be constrained to lie within the
cylinder $x^2+y^2\le1$ since $\Om_\ph\le1$. In the case of the exponential
potential analysed in Ref. \cite{CLW}, one of the differential equations
decoupled, and the dynamics was effectively described by a phase plane with
trajectories bounded by the circle $x^2+y^2=1$. In the present case, however,
no such simplification arises.

\begin{figure}[htp] \centering \leavevmode\epsfysize=8cm
\epsfbox{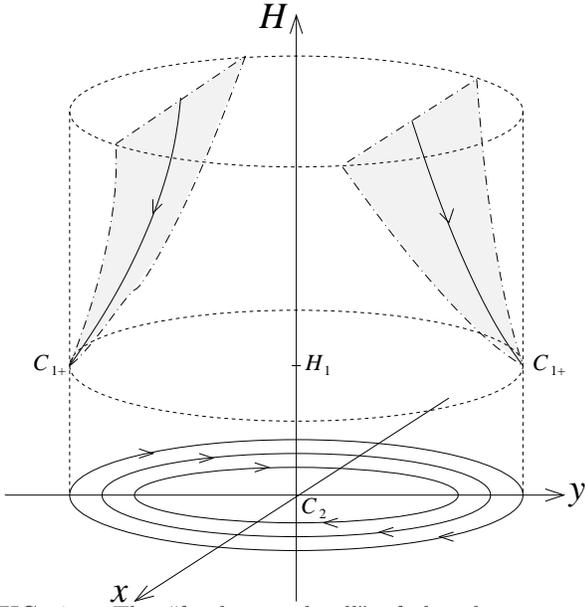} \caption[cell]{\label{cell} The ``fundamental cell'' of
the phase space in terms of the energy density variables. On the
$y^2=H_1^2/H^2$ planes (The shaded planes), only one trajectory is possible as
shown. It corresponds to the scalar field lying on top of the potential hill
all along. On the $H=0$ plane, the trajectories are concentric circles with
centre at the origin.} \end{figure}

The physical region of phase space is further restricted by the requirement
that $y^2\le\frac23m^2/H^2\equiv\Ho^2/H^2$, which is equivalent to $\cos^2\left
(I/2\right)\le1$ in terms of the field variables. For values of $H>\Ho$ each
$H=\const$ slice of the cylinder $x^2+y^2\le1$ is cut off in the $y$-direction
above and below the $y=\pm\Ho/H$ lines. Thus the ``fundamental cell'' of the
phase space can be considered to be a cylinder for $0\le H\le\Ho$, capped by
a horn for $H>\Ho$, which tapers off to a line segment $-1\le x\le1$ on the
$x$-axis as $H\rarr\infty$ (see Fig.\ \ref{cell}). In fact, the phase space
consists of an infinite number of copies of the fundamental cell of
Fig.~\ref{cell} as a result of the periodic structure of the potential. These
cells, $\CC_n$, can be labelled by an integer, $n$, with the variable $I$ lying
in the range $2n\pi\le I<2(n+1)\pi$ for each $n$. For cells with even $n$ the
dynamics is described by (\ref{Hdotxy})--(\ref{mu}) with the upper sign in
(\ref{xdot}) and (\ref{ydot}), while for odd $n$ one must take the lower sign.

Within the horn portion of each phase space cell the motion of most
trajectories is roughly circular in $H=\const$ slices, in a clockwise sense in
even cells, $\CC_{2n}$, and anti-clockwise in odd cells $\CC_{2n+1}$. However,
trajectories can cross from one cell to another along the $y=\pm\Ho/H$
boundaries of their horns, which correspond to the surfaces $I=2n\pi$ in terms
of the field variables. For even cells, $\CC_{2n}$, trajectories join the cell
$\CC_{2n}$ along the $y=\Ho/H$ surface and the cell $\CC_{2n+1}$ along the
$y=-\Ho/H$ surface. Below the $H=\Ho$ plane solutions cannot cross from one
cell to another, but remain confined within the cylinder $x^2+y^2\le1$.

When $H=0$ we see from (\ref{Hdotxy}) - (\ref{mu}) that $\dot{x}=my/\FF$ and
$\dot{y}=-mx/\FF$, so that trajectories which lie in the $H=0$ surface are
purely concentric circles. Since $\dot{a}=0$ in the $H=0$ plane, these do not
represent physically interesting cosmologies, but by continuity the behaviour
of the trajectories just above the plane will be of a spiral nature.

The origin $H=x=y=0$ is in fact the critical point corresponding to $\Co_2$.
The nature of the critical point is altered by the change of variables,
however. In
particular, whereas the eigenvalues for linear perturbations are unchanged,
when higher order corrections are considered the point is no longer always an
asymptotically stable spiral as was the case in Fig.~\ref{rhgm0}.

Asymptotically as $t\rarr\infty$ we have
\bea x=A(t)\sin{mt\over\FF}\ ,\label{Asymx}\\ y=A(t)\cos{mt\over\FF}\ ,
\label{Asymy}\eea
where the amplitude is governed by the equation
\be\label{A2dot}{d\over dt}(A^2)=3HA^2\left(\gm-2\sin^2{mt\over\FF}\right)(1-A
^2)\ .\ee
The nature of $\Co_2$ is now found to depend on $\gm$:

\begin{enumerate} \item If $\gm<1$ we find that over a cycle the average value
of the r.h.s.\ of (\ref{A2dot}) is negative and $A^2$ decreases so that $\Co_2$
is an asymptotically stable spiral. Furthermore to leading order $H\goesas{2
\over3\gm t}$ as $t\rarr\infty$ or $a\propto t^{2/3\gm}$ and $A(t)$ is
given by
\be A(t)=Bt^{(\gm-1)/\gm}\exp\left[-{1\over\gm}\int^\infty_t{{dt'\over t'}
\cos\left({mt'\over\FF}\right)}\right]\ee
with $B$ constant. The late-time attractor has $\Om_\ph=0$ and $\Om_\gm=1$

\item If $\gm>1$ then over a cycle the average value of the r.h.s.\ of
(\ref{A2dot}) is positive and $A^2$ increases until it reaches a limit cycle
$A^2=1$, i.e.\ $x^2+y^2=1$ or $\Om_\ph=1$, $\Om_\gm=0$. In this case
\be A(t)=1-\bar{B}t^{2(1-\gm)}\exp\left[-2\int^\infty_t{{dt'\over t'}\cos
\left({mt'\over\FF}\right)}\right]\ .\ee

\item If $\gm=1$, which corresponds physically to an ordinary matter-dominated
universe, then an intermediate situation obtains. Essentially any of the
concentric circles in the $H=0$ plane of Fig.\ \ref{cell}
can be approached asymptotically
giving a universe for which $\Om_\ph\rarr\al_1$ and $\Om_\ph\rarr1-\al_1$
where $\al_1$ is a constant in the range $0<\al_1<1$, which depends on the
initial conditions and the parameters $m$ and $\FF$. \end{enumerate}

We observe that for all values of $\gm$,
\be\rh_\ph\propto{1\over a^3}\ee
at late times. Since $\rh_\gm\propto a^{-3\gm}$, the three different late time
behaviours can thus be understood as a consequence of the scalar field either
decreasing more rapidly than the barotropic fluid ($\gm<1$), less rapidly
($\gm>1$), or at the same rate ($\gm=1$). The scalar field thus eventually
dominates if $\gm>1$, while the barotropic fluid dominates if $\gm<1$. In
the interesting critical case of dust filled models ($\gm=1$) both the scalar
field and ordinary matter are of cosmologically significant density at late
times.

One simplification that is often made in studying quintessence models
is to assume that at late times the quintessence field obeys an equation
of state
\be P_\ph=\left(\gm_\ph-1\right)\rh_\ph
\ee
with $\gm_\ph$ effectively constant. While such an assumption is justified
in the case of models with a slowly varying scalar field, it does not apply
in the present case. In particular, since the effective barotropic index of
the scalar field is \cite{WCL}
\be \gm_\ph={2x^2\over x^2+y^2}, \ee
we see that $\gm_\ph\simeq\sin^2\left(mt/\FF\right)$ at late times, so that
it remains truly variable, varying from $0$ to $2$ over each cycle. It follows
from (\ref{xdot})--(\ref{Omphidef}) that the scalar energy density parameter
obeys the equation
\be {\dot\Om}_\ph=3H\Om_\ph\left(1-\Om_\ph\right)\left(\gm-\gm_\ph\right)
\label{Omphidot}\ee
so that ${\dot\Om}_\ph\rarr0$ as $t\rarr\infty$, which accords with the
late time properties of the solutions observed above.

\section{Numerical Integration}

We will now consider the extent to which models based on the PNGB potential are
constrained by the latest observational evidence. This work extends the studies
previously undertaken by various authors \cite{VL,FW,CDF,FW2}. In the most
recent analysis, Frieman and Waga have compared the constraints imposed by the
high redshift supernovae luminosity distance on the one hand and gravitational
lensing bounds on the other. The two measures provide tests which are
potentially in opposition.
Here we will perform a similar analysis for the PNGB models, but also taking
into account the possibility of luminosity evolution which has not been
considered in previous studies \cite{FW,FW2}. We are therefore considering the
model of the previous section with $\gm=1$.

To proceed it is necessary to integrate the equations numerically. To do this
we introduce the dimensionless variables
\bea\label{u}u&=&{J\over H_0}={\kp\dot{\ph}\over H_0}\ ,\\ \label{v}v&=&\Om_
{m0}^{1/3}(1+z)\ ,\\ \label{w}w&=&I={\kp\ph\over\FF}\ .\eea
where $\Om_{m0}$ is the fractional energy density of matter at the present
epoch, $t_0$. The dynamical system then becomes
\bea u'&=&-3{H\over H_0}u+{m^2\over\FF H_0^2}\sin w\ ,\\ v'&=&-v{H\over H_0}\ ,
\\ w'&=&{u\over\FF},\eea
where the Hubble parameter is defined implicitly according to
\be\label{Fcuvw}{H\over H_0}\equiv\left[v^3+{1\over6}u^2+{m^2\over3H_0^2}(
\cos w+1)\right]^{1\over2}\ ,\ee
and prime denotes a derivative with respect to the dimensionless time parameter
$\tau\equiv H_0t$. Only the variable $v$ differs from those used by Frieman
and Waga \cite{FW}. Our reason for making the choice (\ref{v}) is that
it allows us to integrate the Friedman equation directly rather than a second
order equation (\ref{HdotIJ}) which follows from the other equations by virtue
of the Bianchi identity. This may possibly lead to better numerical stability
since it is not necessary to implement the Friedmann constraint separately.

We begin the integration at initial values of $u$, $v$, and $w$ chosen to
correspond to initial conditions expected in the early matter-dominated era.
The integration proceeds then until the r.h.s.\ of (\ref{Fcuvw}) is equal to 1,
thereby determining the value of the present epoch, $t_0$, to be the time at
which $H=H_0$. We are then also able to determine $\Om_{m0}$, since according
to (\ref{v})
\be\Om_{m0}=v^3(t_0)\ .\ee

The choice of appropriate initial conditions has been previously discussed
\cite{CDF,VL,FW}.
In particular, since the Hubble parameter is large at early times, it
effectively acts as a damping term in Eq.~(\ref{Hdot}), driving the scalar
field to a state with $\dot{\ph}=0$ initially, i.e.\ $u=0$. We take $v=1101$
initially, which in view of relation (\ref{v}) and the fact that $\Om_{m0}
\goesas0.1-1$ corresponds to the early matter dominated era $1100\le z\lsim
3000$. Results of the integration do not change significantly if $v$ is
altered to values within the same order of magnitude.

The initial value of the scalar field variable, $w_i\equiv w(t_i)$, can lead
to some variability in predicted cosmological parameters at the present epoch.
A few different values of $w(t_i)$ have been considered by different authors
\cite{CDF,VL,FW}. However, the only systematic studies of bounds in the
$M$,$\ff$
parameter space have been performed \cite{FW,FW2} for one particular initial
value, $w_i=1.5$. One must bear in mind that such bounds are also dependent
on $w_i$, the value of which is not greatly restricted. Given that we are
starting with $u(t_i)\simeq0$, so that the kinetic energy of the scalar field
is initially negligible, the only physical restriction on the value of $w_i$
comes from the requirement that the scalar field should be sufficiently far
from the minimum of the potential, $V(\ph)$, that $\Om_\ph(t_i)$ is small.
Thus will ensure that $w_i$ is consistent with a scalar field that has emerged
from the radiation dominated era with $\Om_\ph$ sufficiently small that is
consistent with bounds set by primordial nucleosynthesis, and by structure
formation models. This still leaves considerable latitude for the choice
of $w_i$, however.

\begin{figure*}[htp] \centering\leavevmode\vskip2pt
\epsfysize=8cm \epsfbox{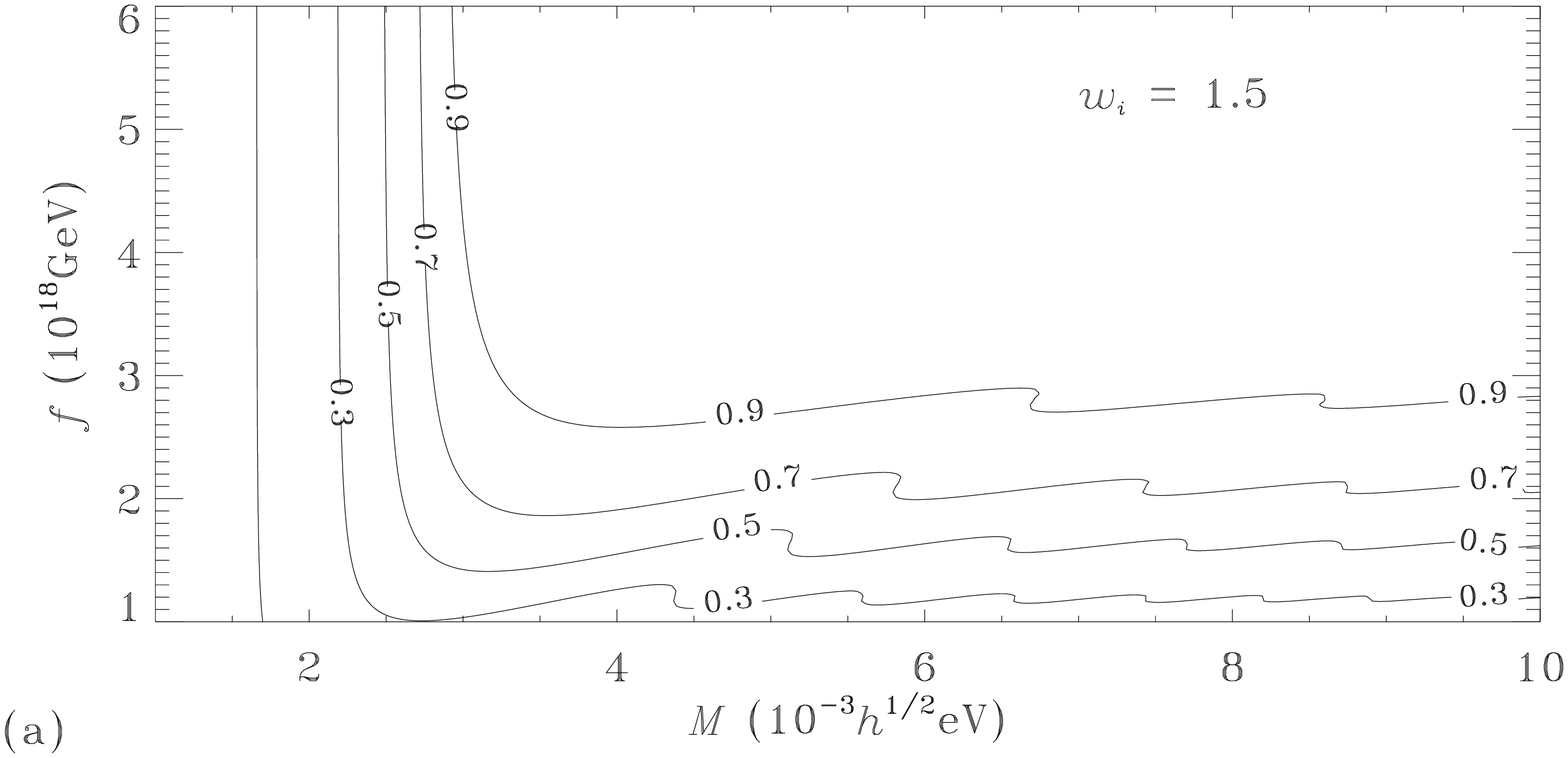}\vskip2pt
\epsfysize=8cm \epsfbox{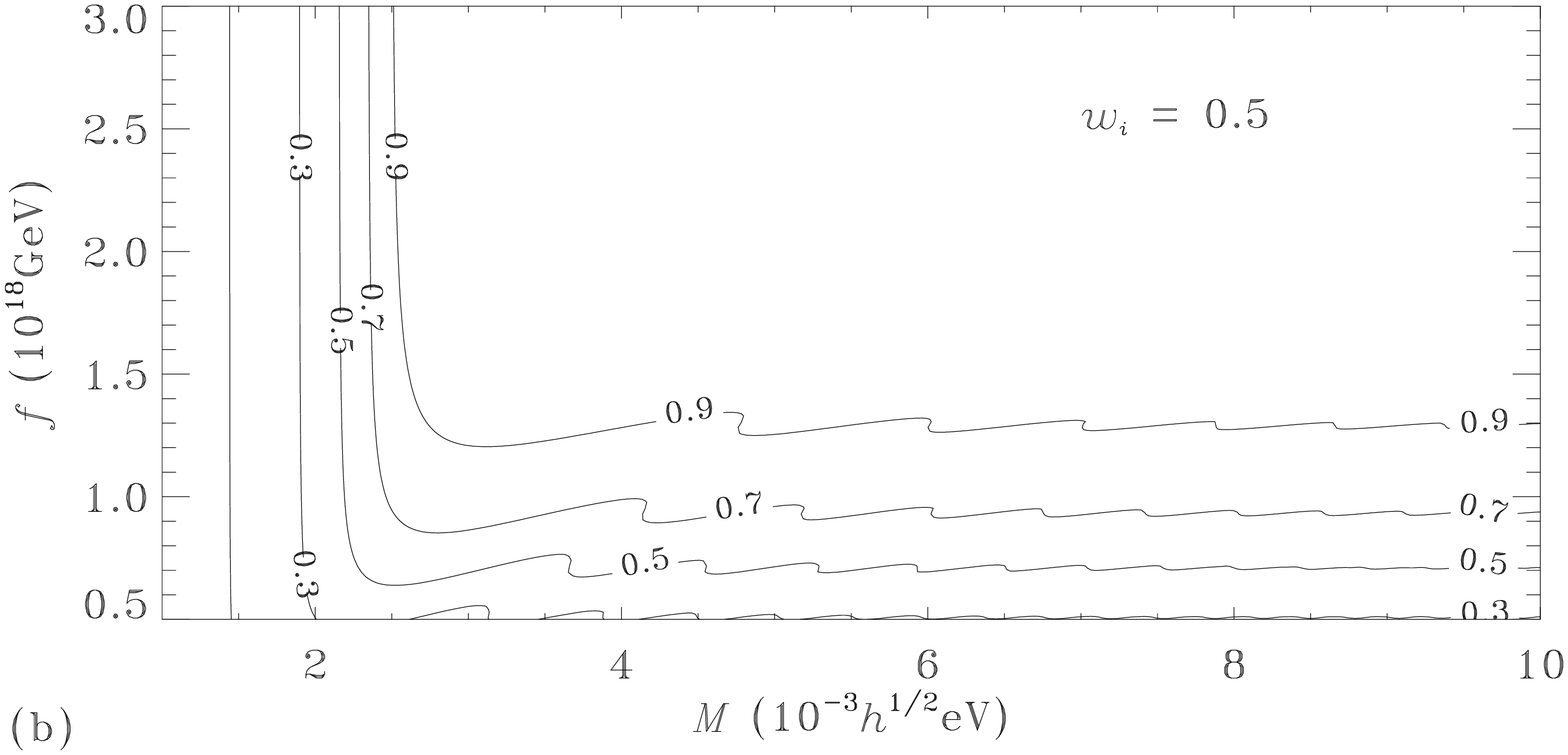}
\caption[Omphi0]{\label{Omphi0} Contours of $\Omp$ in the $M$,$\ff$
parameter space for two choices of initial values: (a) $w_i=1.5$; (b)
$w_i=0.5$} \end{figure*}
\begin{figure*}[htp] \centering\leavevmode\vskip2pt
\epsfysize=8cm \epsfbox{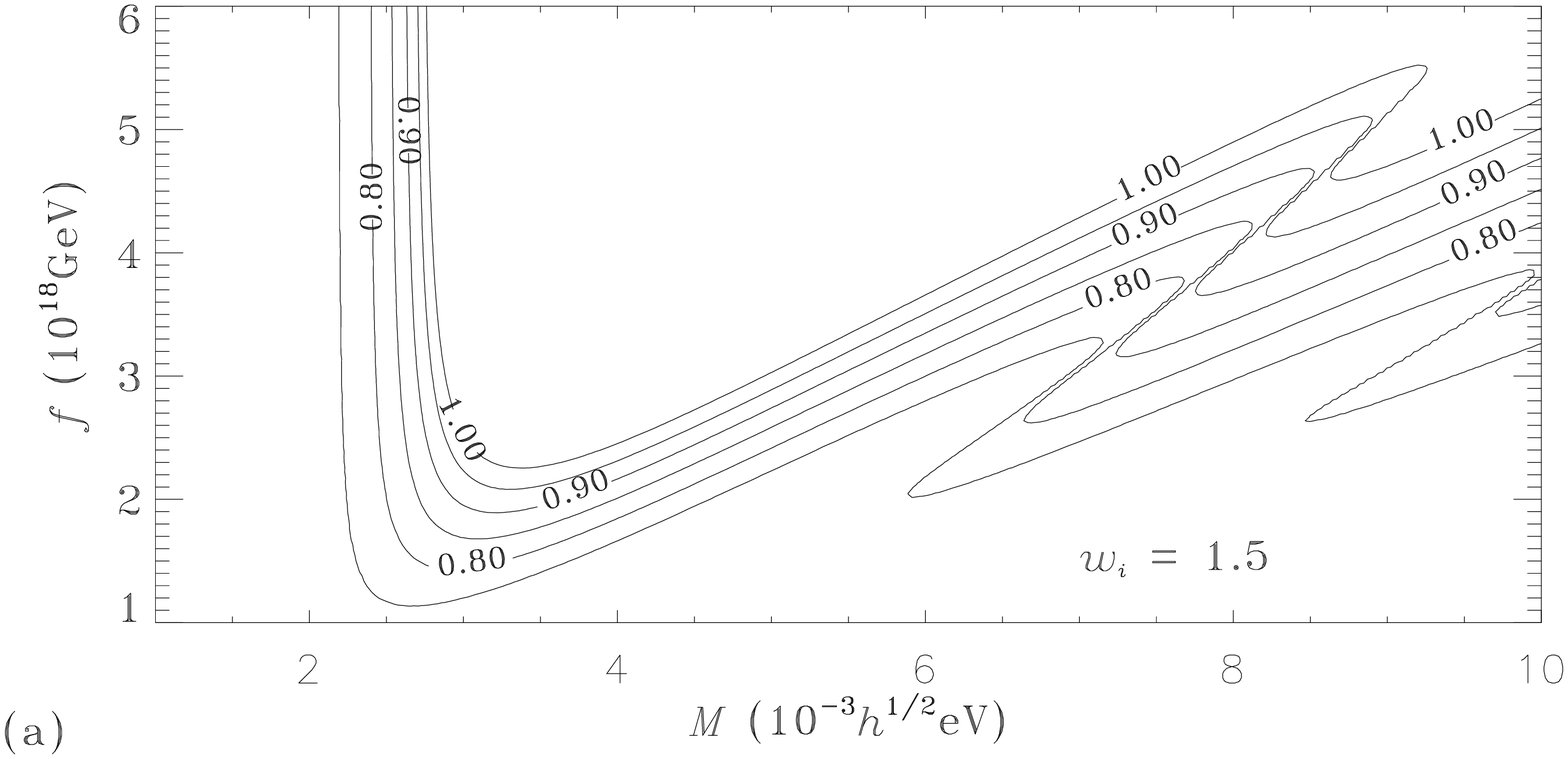}\vskip2pt
\epsfysize=8cm \epsfbox{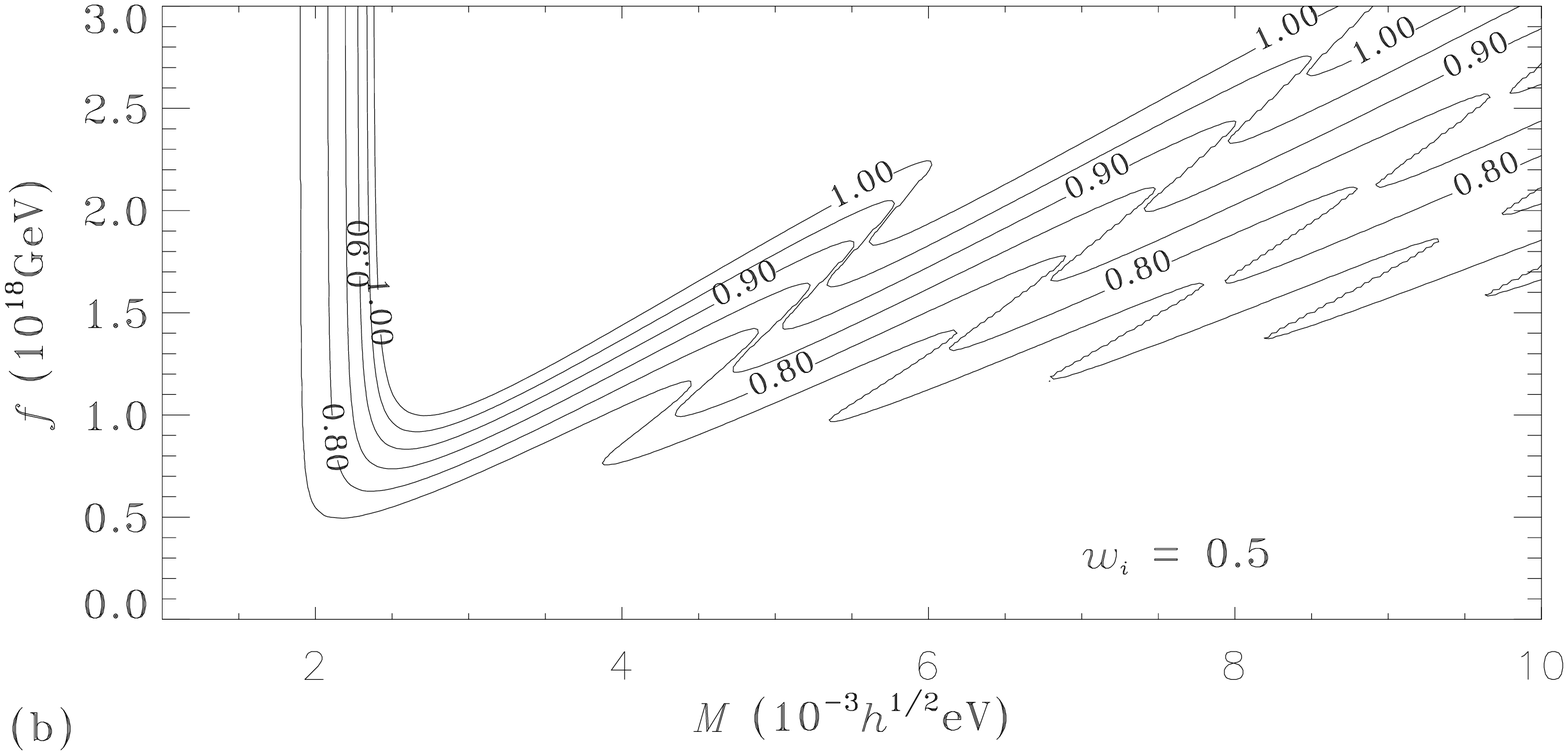}
\caption[H0t0]{\label{H0t0} Contours of $H_0t_0$ in the $M$,$\ff$
parameter space for two choices of initial values: (a) $w_i=1.5$; (b)
$w_i=0.5$} \end{figure*}
\begin{figure*}[htp] \centering\leavevmode\vskip2pt
\epsfysize=8cm \epsfbox{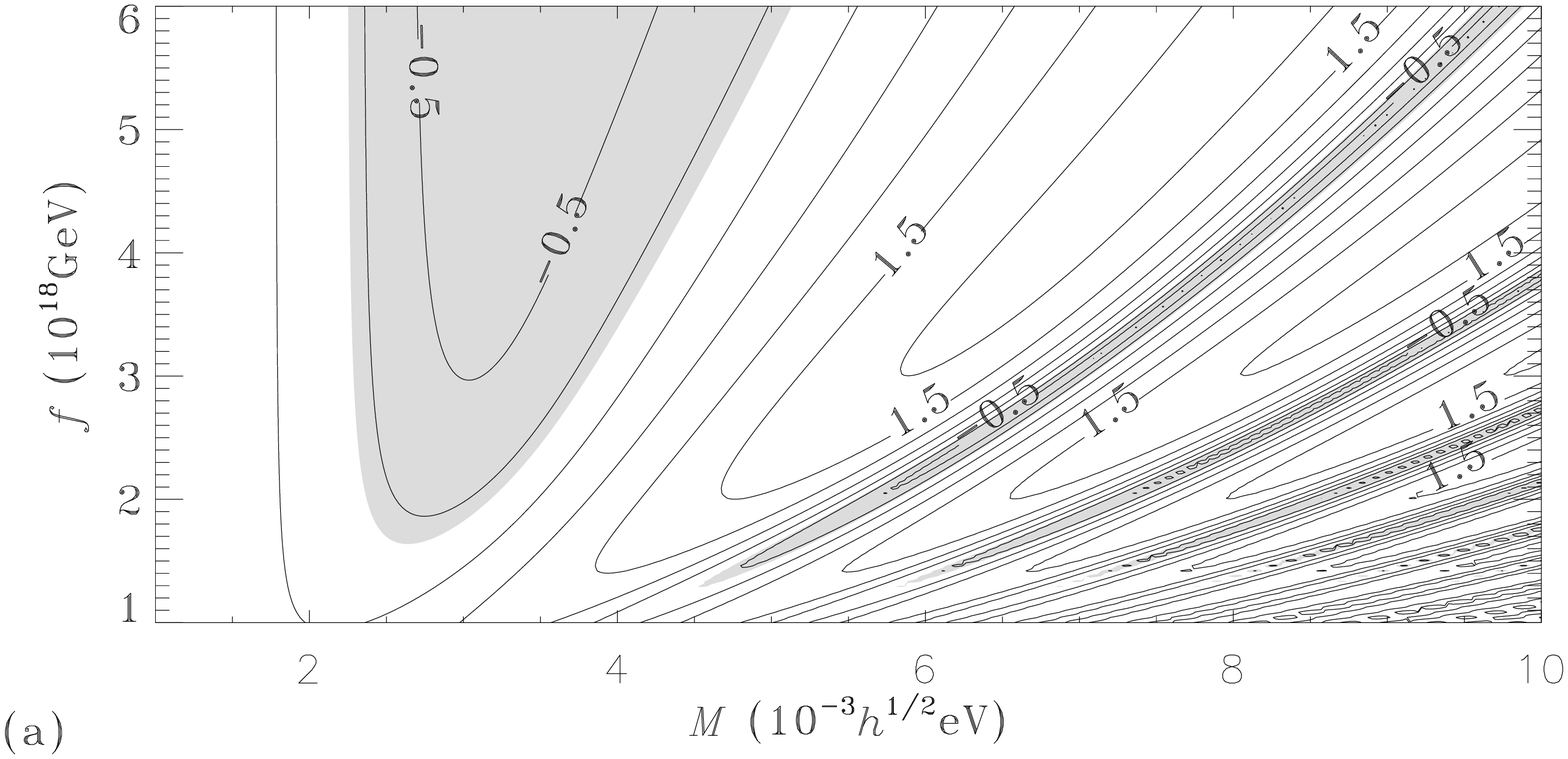}\vskip2pt
\epsfysize=8cm \epsfbox{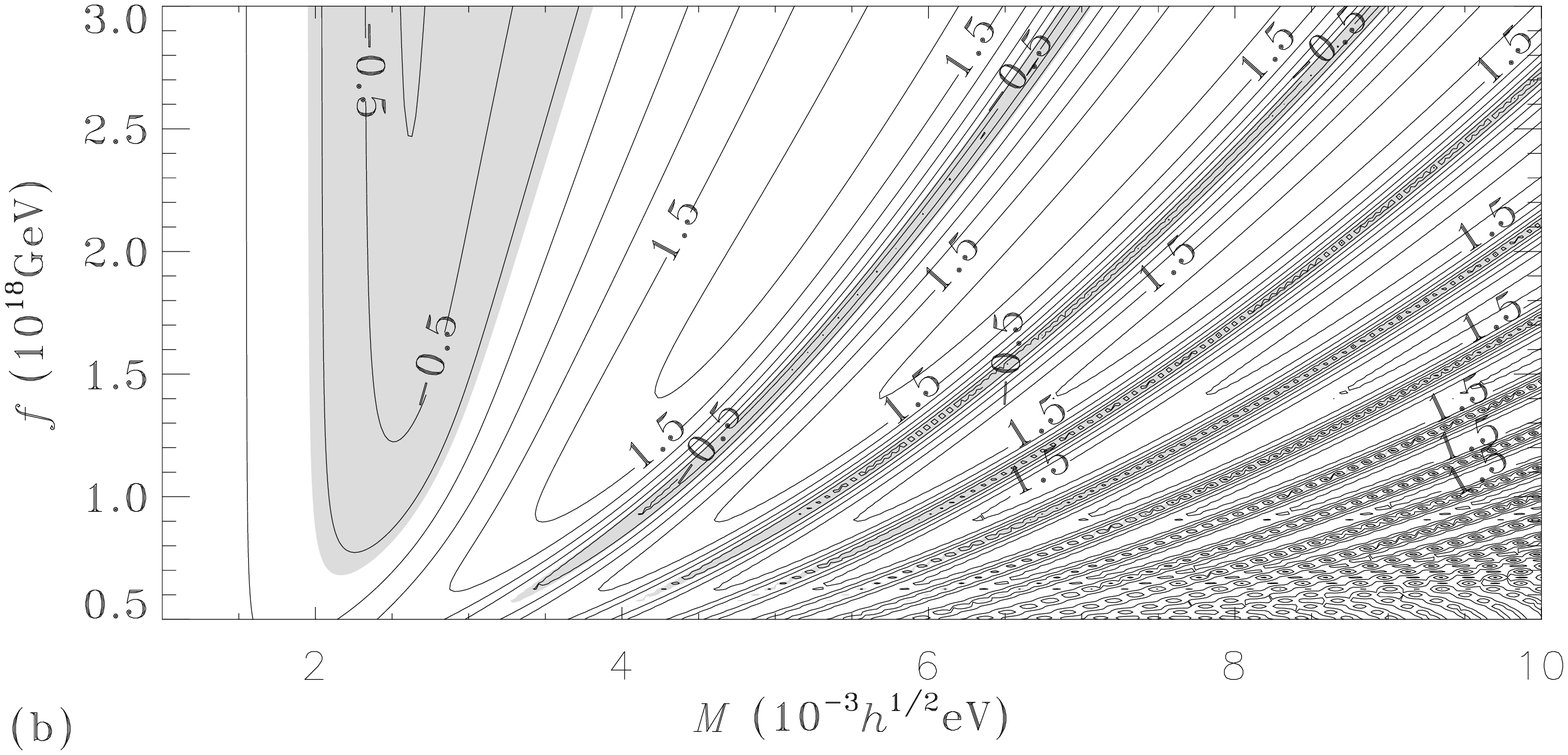}
\caption[qn]{\label{qn} Contours of $\qn$ in the $M$,$\ff$
parameter space for two choices of initial values: (a) $w_i=1.5$; (b)
$w_i=0.5$. Values $\qn<0$, corresponding to a universe whose expansion is
accelerating at the present epoch, are shaded.} \end{figure*}

In Figs.\ \ref{Omphi0} and \ref{H0t0} we display contour plots of
$\Omp$ and $H_0t_0$ in the $M$,$\ff$ parameter space for
two values $w_i=1.5$ and $w_i=0.5$. Similar figures have been given
by Frieman and Waga \cite{FW} in the $w_i=1.5$ case, although our resolution
is somewhat better. As $w_i$ decreases the contour plots do not change
significantly in terms of their overall features, but contours with equivalent
values shift to lower values of
the $\ff$ parameter. For example, for large values of $M$ the $\Omp=0.7$
contour lies at a value $\ff\simeq2.05\times10^{18}\w{GeV}$ if $w_i=1.5$,
while the same contour lies at $\ff\simeq0.94\times10^{18}\w{GeV}$
if $w_i=0.5$.

The other principal feature of the plots \ref{Omphi0} and \ref{H0t0}, which
was not commented on in Ref.\ \cite{FW}, is the wave-like properties of the
contours at larger values of $M$. These features can be readily understood
by considering the corresponding plots of the deceleration parameter, $\qn$,
which is defined in terms of $u(t_0)$, $v(t_0)$, and $w(t_0)$ by
\bea\qn&=&{3\over2}v^3(t_0)+{1\over2}u^2(t_0)-1\ ,\nonumber\\
&=&\frac12\left(1+3\Omp-6{y_0}^2\right)\ .\eea
We display contour plots of $\qn$ in the $M$,$\ff$ parameter space in Fig.\
\ref{qn}. Essentially, as $M$ increases for roughly fixed $\ff$ for
sufficiently large $M$, the value of $\qn$ oscillates over negative and
positive values, from a minimum of $\qn=\frac12\left(1-3\Omp\right)$ to a
maximum $\qn=\frac12\left(1+3\Omp\right)$ about a mean of $\qn=0.5$. This
corresponds to the scalar field, $\ph$, having undergone more and more
oscillations by the time of the present epoch. The minimum value of $\qn$ is
attained when $\dot\ph=0$ instantaneously, while the maximum value of $\qn$
is attained when $\ph$ is instantaneously passing through the minimum of its
potential.

For smaller values of $M$ to the left of the plots the scalar
field has only relatively recently become dynamical, whereas for larger
values of $M$, the scalar can already have undergone several oscillations
by the time of the present epoch, particularly if $\ff$ is small. This
variation can be understood in terms of the asymptotic period of oscillation
of solutions which approach $\Co_1$, which by Eqs.\ (\ref{Asymx}),
(\ref{Asymy}) is
\be t_a=2\pi\FF/m=2\pi\ff/M^2. \ee
The period $t_a$ is shorter for larger $M$, or for smaller $\ff$.
Since the final $\ff$ values plotted in the $w_i=0.5$ case are a factor of
2 smaller than the $w_i=1.5$ case, this also explains why points with the
same value of $M$ have undergone more oscillations up to the present epoch
for the smaller value of $w_i$.

The value of $H_0t_0$ oscillates as $M$ increases for roughly fixed $\ff$,
according to whether the universe has been accelerating or decelerating in the
most recent past, with more rapid variation for parameter values with shorter
asymptotic periods, $t_a$.

The wiggles in the $\Omp$ contours are a
residual effect of the oscillation of the scalar field as $\Om_\ph$ settles
down to a constant value according to (\ref{Omphidot}). The variation in the
value of $\Omp$ with the value of $w_i$ can be understood from the
fact that for smaller values of $w_i$ the scalar field begins its evolution
in the matter-dominated era closer to the critical points, $\Co_{1+}$,
corresponding to the maximum of the potential, $V(\ph)$. For fixed $M$ and
$\ff$ the period of quasi-inflationary expansion is therefore longer, and the
present value of $\Omp$ larger.

\section{Constraints from High-redshift type IA Supernovae}

Empirical calibration of the light curve - luminosity relationship of type Ia
supernovae provides absolute magnitudes that can be used as distance
indicators. Since the luminosity distance of a light source, $d_L$, is
defined by
\be{H_0d_L\over c}=(z+1)\int^{\tau_0}_\tau(z+1)\,d\tau, \label{lumdist}\ee
with $\tau=H_0t$, it is convenient to define an additional variable
\be r=-\int^\tau_{\tau_0} v\,d\tau\ ,\ee
which is proportional to the comoving coordinate distance
\be r\ns{comoving}={r\over H_0a(t_0)v(t_0)}\,. \ee
We then adjoin a differential equation
\be r'=-v\ee
to the differential equations (\ref{u})-(\ref{w}) when performing the numerical
integration. In terms of $r$ and $v$, the luminosity distance is then
determined according to
\be\label{dL}{H_0d_L\over c}={v\over\Om_{m0}^{2/3}}(r-r(t_0))\ ,\ee
and can be used in the appropriate distance modulus to compare with the
supernovae data.

Frieman and Waga \cite{FW2} have recently considered constraints on PNGB models
from supernovae data (without source evolution) using the data set of Riess
\etal\ \cite{R98}. We will perform a similar analysis, but we will make
use of the largest available data set, namely the 60 supernovae published
by Perlmutter \etal\ \cite{P98} (hereafter P98). Of these, 18 low redshift SNe
Ia were discovered and measured in the Cal\'an-Tololo survey \cite{H96c}, and
the Supernova Cosmology Project discovered 42 new SNe Ia at redshifts between
0.17 and 0.83. The peak magnitudes of the supernovae are corrected using the
``stretch factor'' light curve fitting method \cite{P98,P97}. This method is
based on fitting a time-stretched version of a single standard template to the
observed light curve. The stretch factor is then used to estimate the absolute
magnitude.

\subsection{Models without evolution}

\begin{figure*}[htp] \centering\leavevmode
\epsfysize=8cm \epsfbox{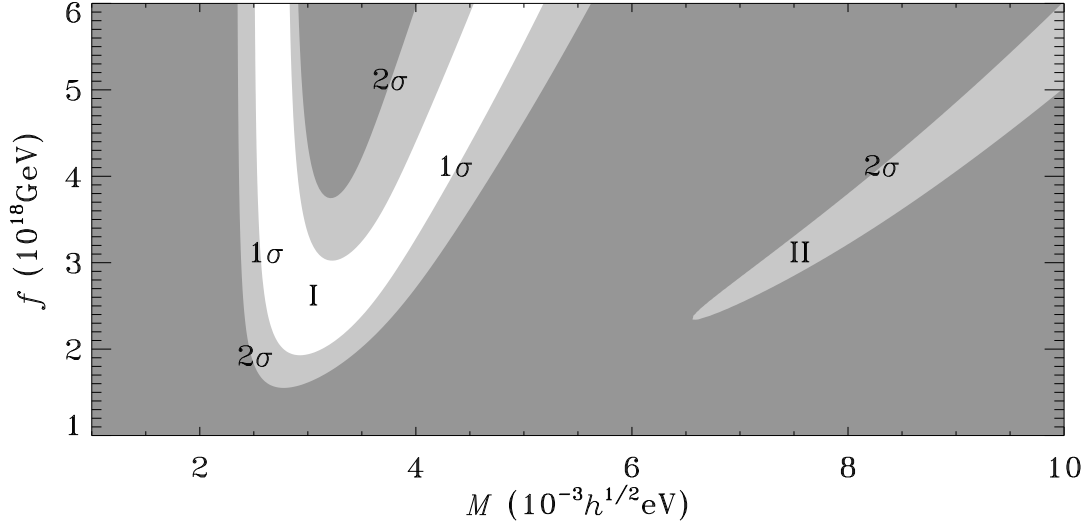}
\caption[SnIabeta0]{\label{SnIabeta0} Confidence limits on $M$,$\ff$
parameter values, with $w_i=1.5$ and no evolution of sources, for the 60
supernovae Ia in the P98 dataset.
Parameter values excluded at the 95.4\% level are darkly shaded, while those
excluded at the 68.3\% level are lightly shaded.} \end{figure*}

We use the stretch-luminosity corrected effective $B$-band peak magnitude in
Table 1 of P98 as the absolute magnitude and denote it as $\meff B$. The
distant modulus of a supernova is defined as $\meff B+\MM$, where $\MM$ is
the fiducial absolute magnitude which has not been given in the literatures. A
fitting method to obtain $\MM$, using only the 18 Cal\'an/Tololo supernovae,
can be found in \cite{H95,H96}. Perlmutter \etal \cite{P98} calculate the
confidence regions by simply integrating over $\MM$. In this paper, we will
perform an analytic marginalization over $\MM$ and obtain the marginal
likelihood.

Similarly to the $\chi^2$ statistic used by Riess \etal\ \cite{R98}, we define
the quadratic form
\be\label{chi2}\chi^2(M,f)=\sum^{60}_{i=1}{(\meff{B,i}+\MM-\mu_i)^2\over
\si_i^2}\ee
where $\mu_i(z_i;H_0,M,\ff)$ is the predicted distance modulus for model
parameters $(M,\ff)$, and
\be
\si_i^2=\si^2_{\meff{B,i}}+\left({5\log\e\over z_i}\si_{z,i}\right)^2\ .
\ee
The predicted distance modulus is
\be\mu_i=5\log{d_L(z_i)}+25\ ,\ee
if the luminosity distance, $d_L$, as defined by (\ref{lumdist}) or (\ref{dL}),
is given in units of Megaparsecs.

In order to perform analytic marginalization over $H_0$ as well as over $\MM$
we separate out the $H_0$ and $\MM$ dependence from $\mu_i$ into a quantity,
$\nu$, which we define by
\be\label{muiM0}\mu_i-\MM=g_i+\nu\ee
where $g_i(z_i;M,\ff)$ depends implicitly only on $M$ and $\ff$. We then
follow the statistical procedures adopted by Drell, Loredo and Wasserman
\cite{DLW,DLWnote} and marginalize over $\nu$ using a flat prior that
is bounded over some range $\Delta\nu$.

The marginal likelihood is
\be\label{calL}{\cal L}(M,\ff)={1\over\Delta\nu}\int{d\nu\;\e^{-\chi^2/2}}={s
\sqrt{2\pi}\over\Delta\nu}\e^{-Q/2}\ \ee
where
\be\label{Q1}Q(M,\ff)=\sum^{60}_{i=1}{(\meff{B,i}-g_i-\hat{\nu})^2\over
\si_i^2}\ee
and $\hat{\nu}(M,f)$ is the best-fit value of $\nu$ given $M$ and $\ff$, with
conditional uncertainty $s$. We identify $1/s^2$ as the coefficient of $\nu^2$
in (\ref{chi2}),
\be\label{s}{1\over s^2}=\sum^{60}_{i=1}{1\over\si_i^2}\ ,\ee
and $\hat{\nu}$ as $s^2$ times the coefficient of $-\nu$, viz.\
\be\label{gmhat}\hat{\nu}(M,f)=s^2\sum^{60}_{i=1}{\meff{B,i}-g_i\over\
\si_i^2}\ .\ee

The quadratic form (\ref{Q1}) is what one would obtain by calculating the
``maximum likelihood'' for $M$ and $\ff$. Since $s$ is independent of $M$ and
$\ff$, it follows from Eq.~(\ref{calL}) that the marginal likelihood is
proportional to the maximum likelihood in this case.

Fig.~\ref{SnIabeta0} shows the 68.3\% and 95.4\% joint credible regions for $M$
and $f$, and is a direct analogue of Fig.\ 1 of Ref.\ \cite{FW2}, where a
similar analysis was performed on 37 supernovae given in R98. The position
of the region of parameters which are included at both the 68.3\% and
95.4\% confidence levels is broadly similar to that obtained from the R98
data \cite{FW2}. Although Frieman and Waga \cite{FW2} did not include the
parameter region $M>0.004h\w{eV}$, we have redone their analysis on the R98
data and find that the parameter values to the right of Fig.~\ref{SnIabeta0}
which are admitted at the 95.4\% level but excluded at the 68.3\% level for the
P98 dataset, (labelled region II in Fig.~\ref{SnIabeta0}), are in fact excluded
also at the 95.4\% confidence level if the 37 supernovae of the R98 dataset
are used. It is possible that this discrepancy
has its origin in the different techniques used by Riess \etal\ \cite{R98} to
determine the distance moduli. Possible systematic discrepancies in the
``stretch factor'' method of P98 versus the ``multi-colour light curve'' and
``template fitting methods'' of R98 have been discussed in some detail in
Ref.\ \cite{DLW}.

The importance of the $2\si$ included parameter region to right of Fig.\
\ref{SnIabeta0} diminishes, however, if one compares it with Figs.\
\ref{Omphi0} and \ref{H0t0}, since it largely corresponds to parameter values
with $\Omp\gsim0.9$, which can be discounted by dynamical measurements
of $\Om_{m0}$, where $\Om_{m0}+\Omp=1$. Furthermore, the few allowed
values below the $\Omp=0.9$ contour in this part of the parameter space
have unacceptably small values for the age of the Universe, $H_0t_0$.

The parameter region $0.002h\w{eV}<M<0.003h\w{eV}$ which from Fig.\
\ref{SnIabeta0} is admitted at both the 68.3\% and 95.4\% levels, (labelled
region I), by contrast, corresponds to acceptable values of both $\Omp$ and
$H_0t_0$. Comparing with Fig.~\ref{qn}, we see that this region has
$-0.1\lsim\qn\lsim-0.6$, corresponding to a Universe with a scalar field
still in an early stage of rolling down the potential $V(\ph)$. The label I
is thus indicative of the fact that the scalar field is rolling down the
potential for the first time (from left to right), while in region II the
scalar field is rolling down the potential for the second time (from right to
left). In region II $\qn$ is positive -- however, it corresponds to
parameter values for which there would have been a cosmological acceleration
at modest redshifts in the past, e.g., at $z\goesas0.2$, well within the
range of the current supernovae dataset.

Since conclusions regarding statistically preferred regions of the parameter
space can change if evolution of the sources occurs, we think it is important
that this possibility is also examined, as we will now do.

\subsection{Models with evolution}

\begin{figure*}[htp] \centering\leavevmode\vskip-32pt
\epsfysize=8cm \epsfbox{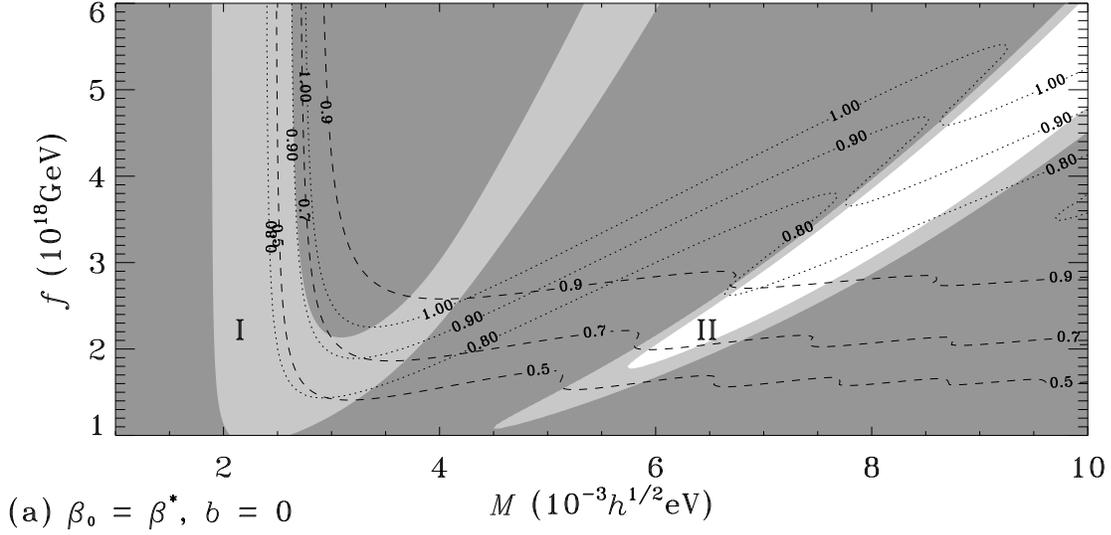}\vskip-10pt
\epsfysize=8cm \epsfbox{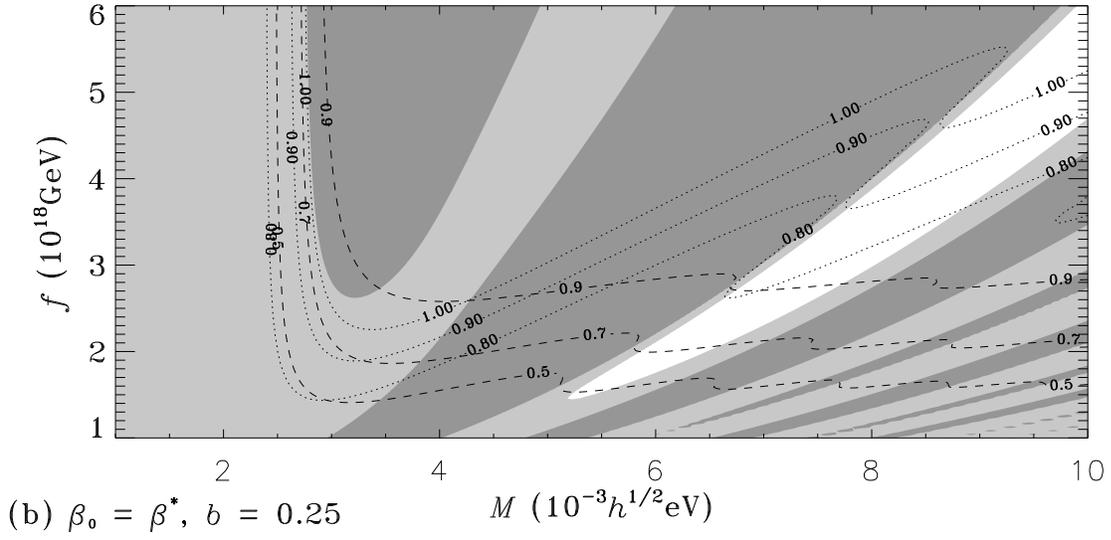}\vskip-10pt
\epsfysize=8cm \epsfbox{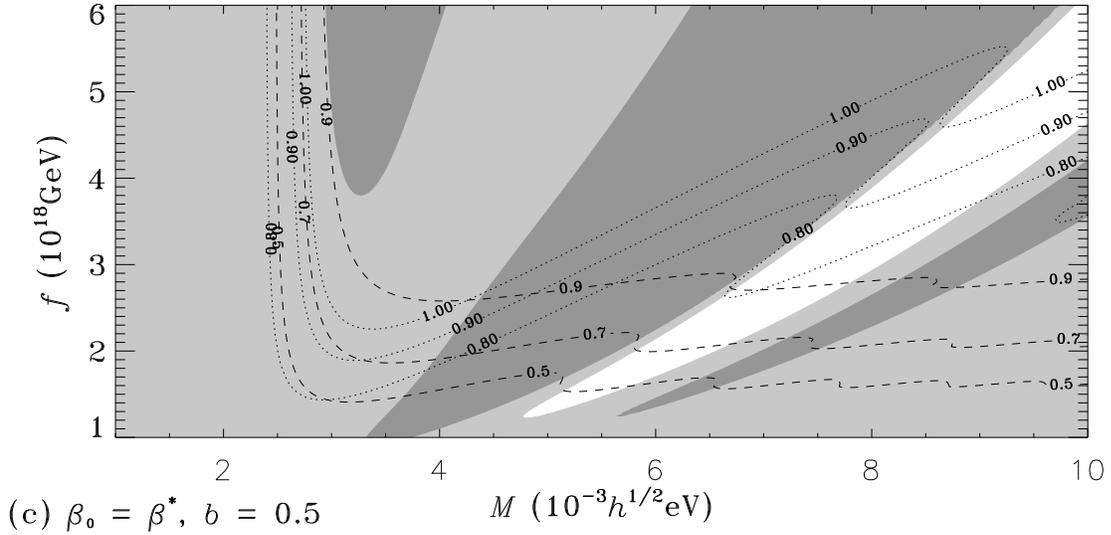}
\caption[SnIa-evolv1]{\label{SnIa-evolv1} Confidence limits on $M$,$\ff$
parameter values in the best-fit $\beta_0=0.414$ slice of the
$(M,\ff,\beta_0)$ parameter space, with $w_i=1.5$, for the 60
supernovae Ia in the P98 dataset.
Parameter values excluded at the 95.4\% level are darkly shaded, while those
excluded at the 68.3\% level are lightly shaded.
For reference, contours of $\Omp$ and $H_0t_0$ are superposed as dashed and
dotted lines respectively.} \end{figure*}

\begin{figure*}[htp] \centering\leavevmode\vskip-32pt
\epsfysize=8cm \epsfbox{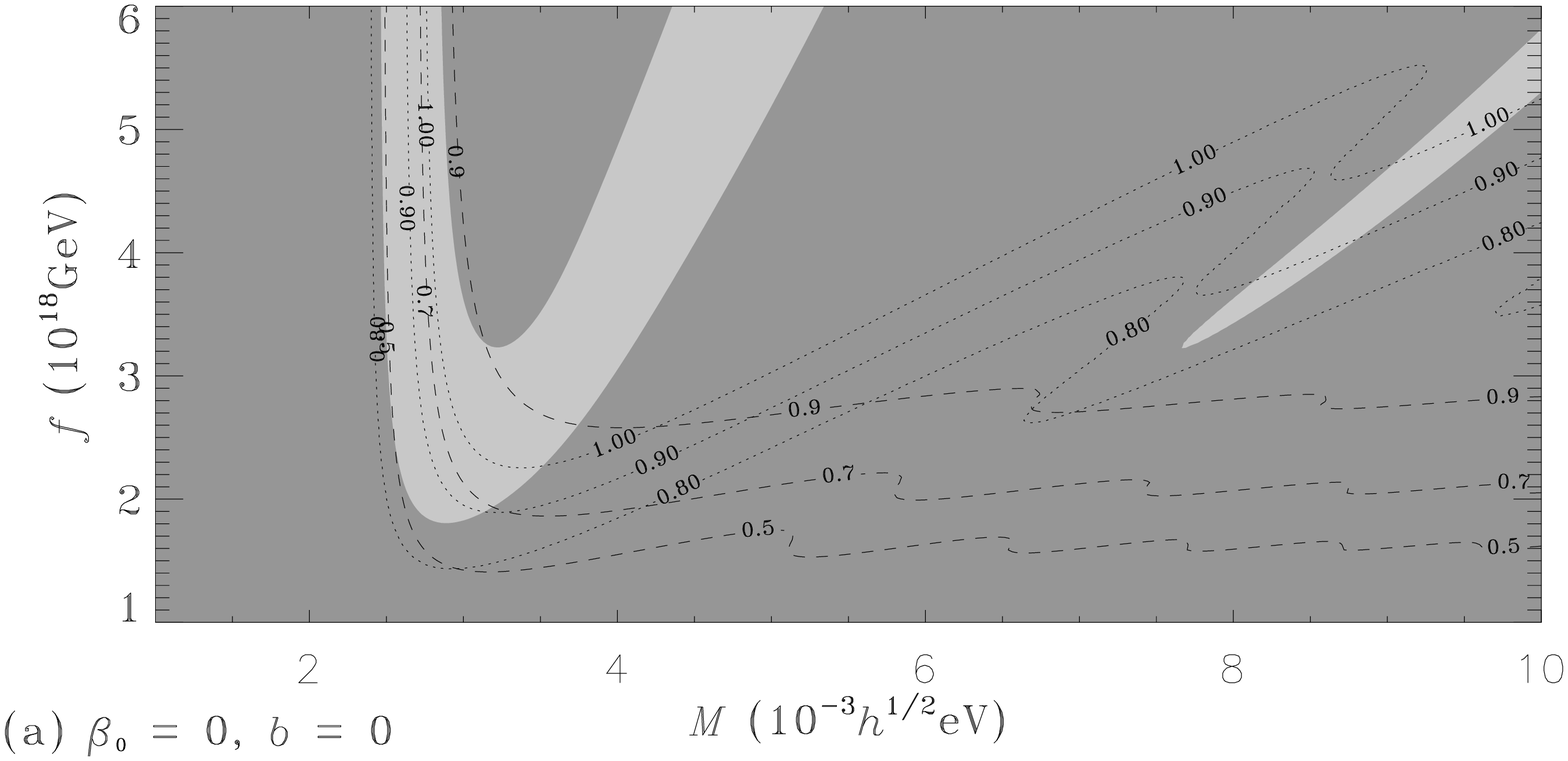}\vskip-10pt
\epsfysize=8cm \epsfbox{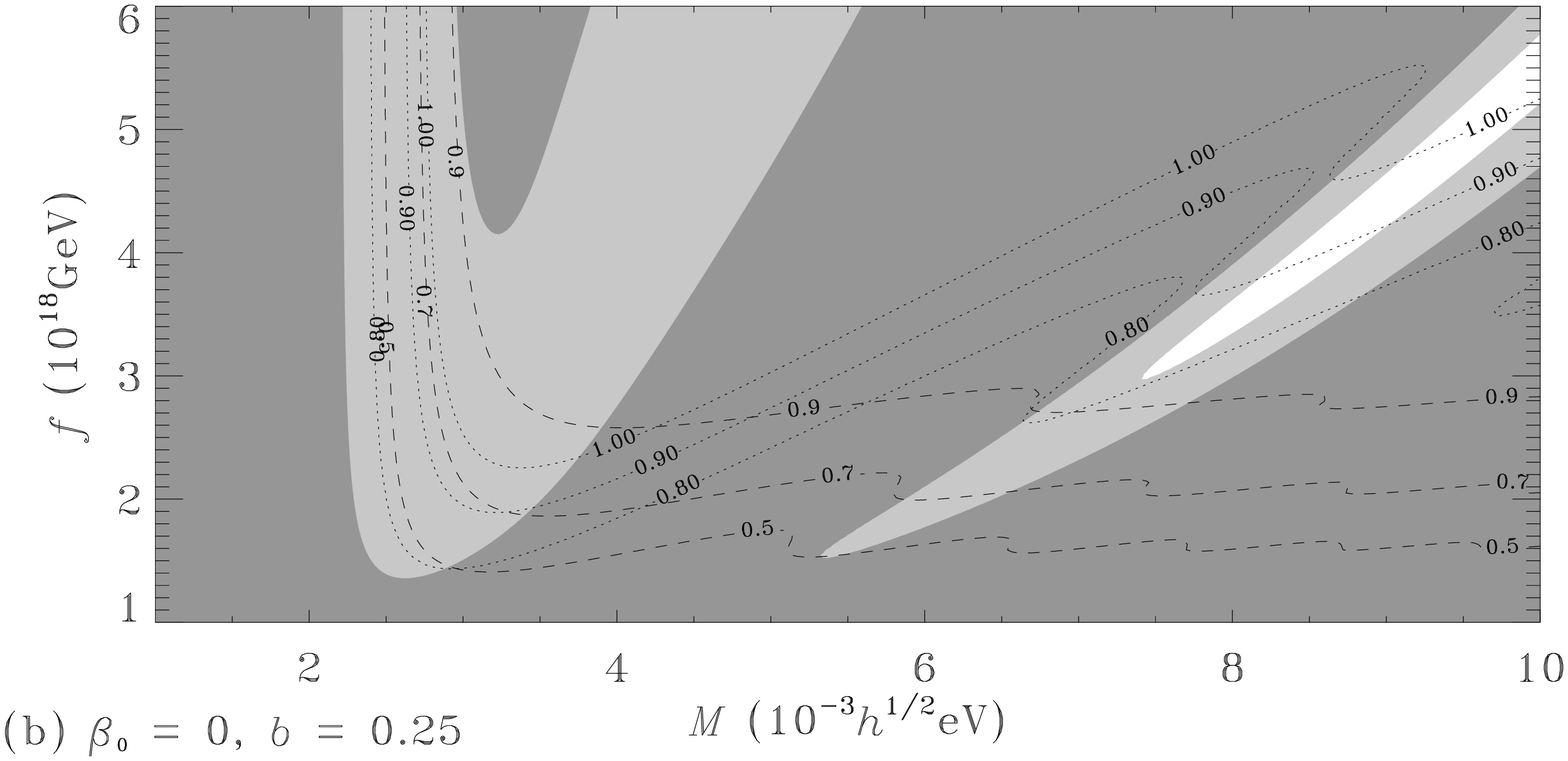}\vskip-10pt
\epsfysize=8cm \epsfbox{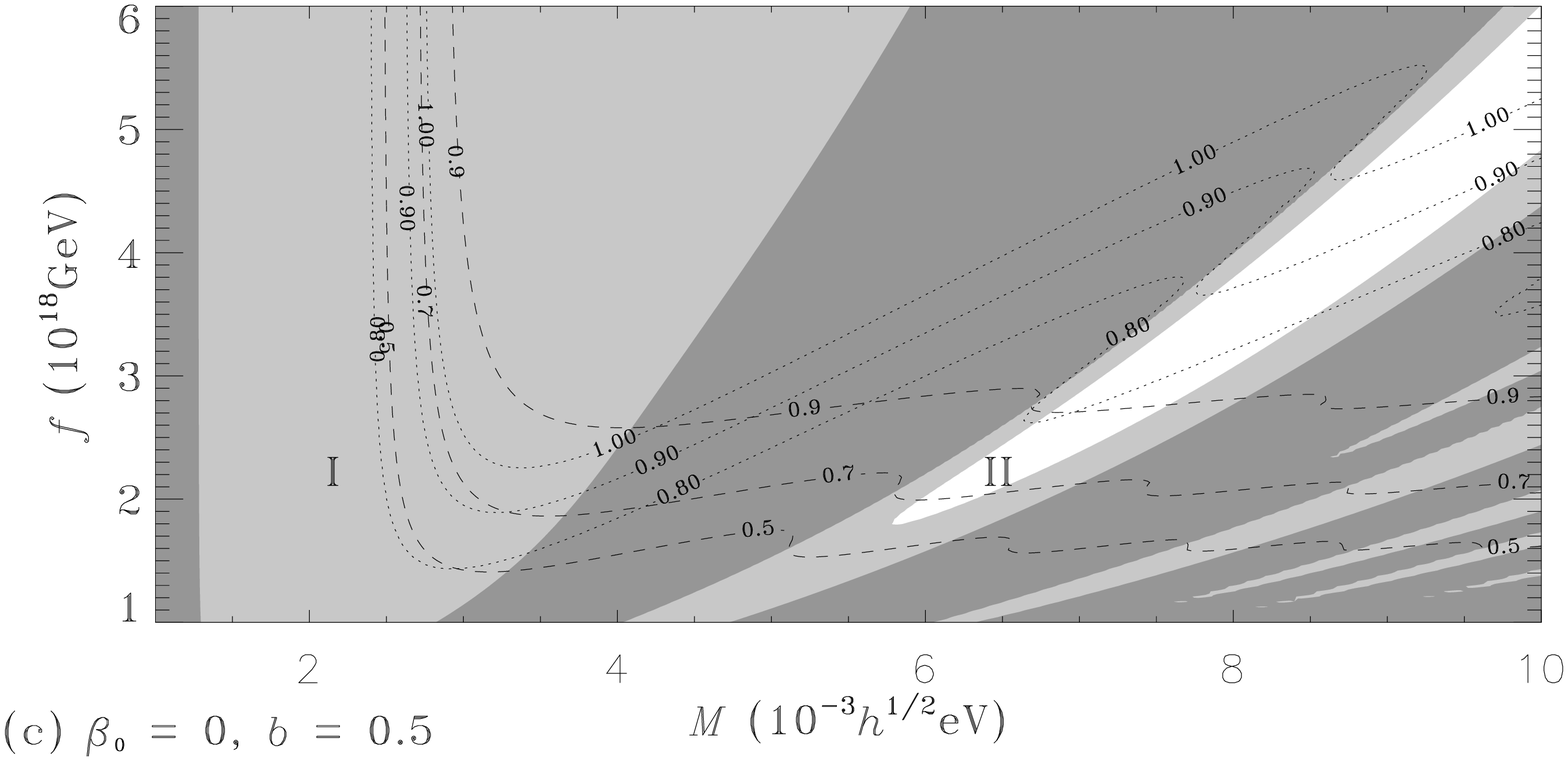}
\caption[SnIa-evolv2]{\label{SnIa-evolv2} Confidence limits on $M$,$\ff$
parameter values in the $\beta_0=0$ slice of the $(M,\ff,\beta_0)$
parameter space relative to a best-fit value $\beta^*\simeq0.414$, for the
60 supernovae Ia in the P98 dataset, with $w_i=1.5$.
Parameter values excluded at the 95.4\% level are darkly shaded, while those
excluded at the 68.3\% level are lightly shaded. For reference, contours of
$\Omp$ and $H_0t_0$ are superposed as dashed and
dotted lines respectively.} \end{figure*}

\begin{figure*}[htp] \centering\leavevmode\vskip-32pt
\epsfysize=8cm \epsfbox{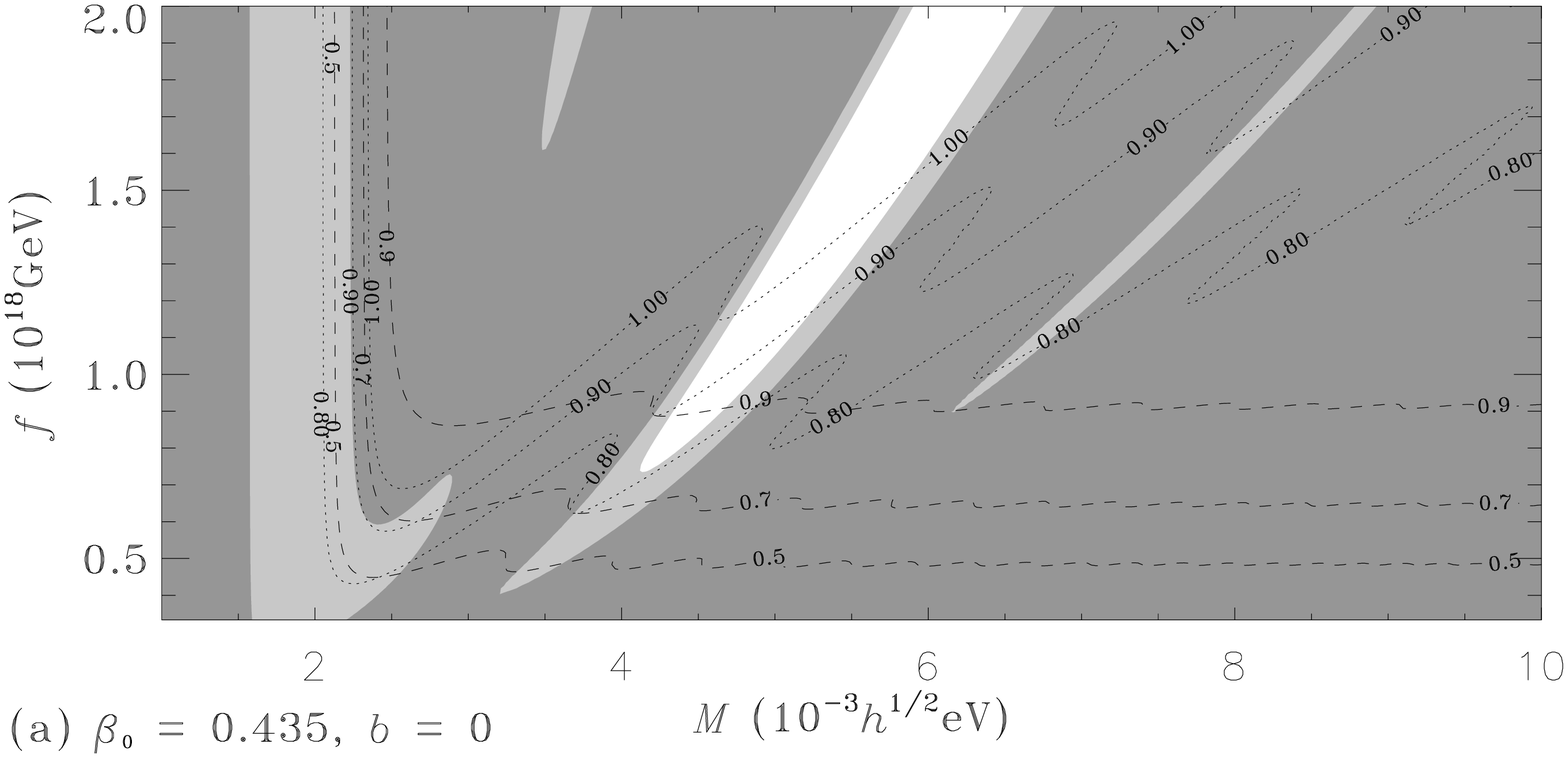}\vskip-10pt
\epsfysize=8cm \epsfbox{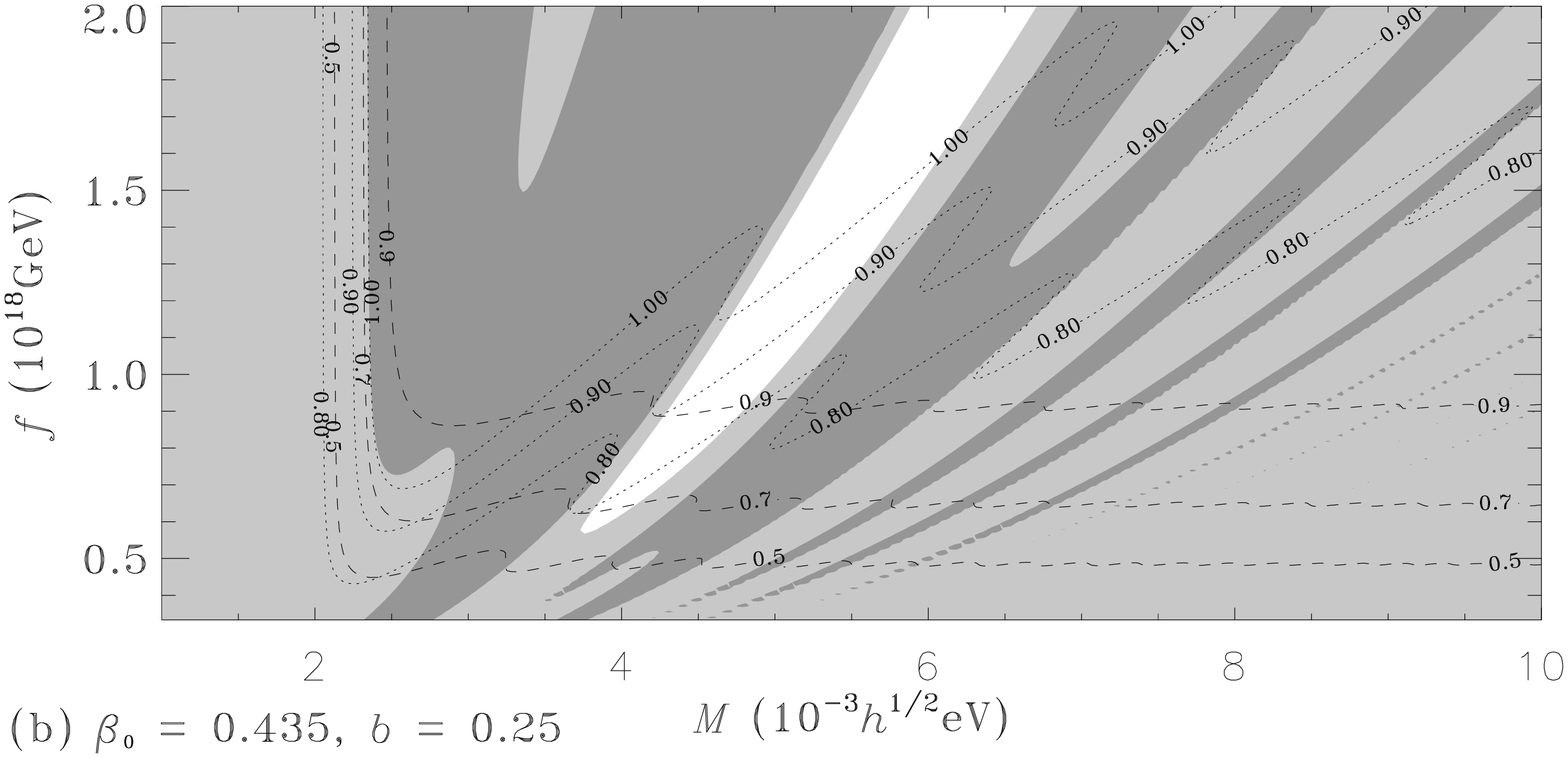}\vskip-10pt
\epsfysize=8cm \epsfbox{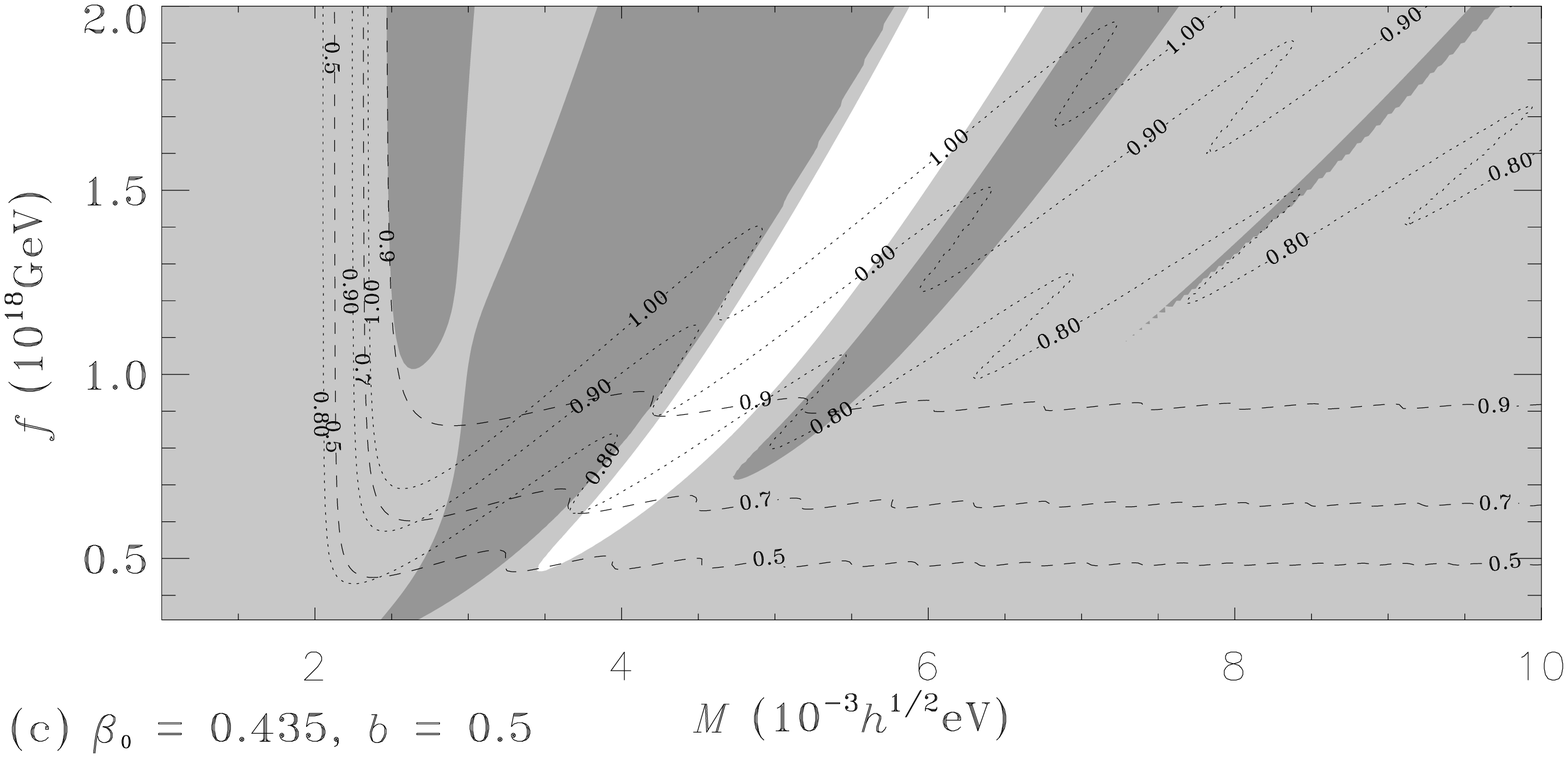}
\caption[SnIa-evolv3]{\label{SnIa-evolv3} Confidence limits on $M$,$\ff$
parameter values in the $\beta_0=0$ slice of the $(M,\ff,\beta_0)$
parameter space relative to a best-fit value $\beta^*\simeq0.435$, for the
60 supernovae Ia in the P98 dataset, with $w_i=0.2$.
Parameter values excluded at the 95.4\% level are darkly shaded, while those
excluded at the 68.3\% level are lightly shaded. For reference, contours of
$\Omp$ and $H_0t_0$ are superposed as dashed and
dotted lines respectively.} \end{figure*}

In view of the recent results of Refs.\ \cite{Rie1,AKN}, supernovae Ia
may exhibit some evolutionary behaviour, at least as far as their rise times
are concerned. A possible evolution in the peak luminosity is therefore a
possibility which must be seriously investigated.

In the absence of a detailed physical model to explain precisely how the
source peak luminosities vary with redshift, one approach is to assume some
particular empirical form for the source evolution, and to examine the
consequences. Such an analysis has been recently preformed by Drell, Loredo
and Wasserman \cite{DLW} in the case of Friedmann-Lema\^{\i}tre models with
constant vacuum energy. We will undertake an equivalent analysis for the
case of PNGB quintessential cosmologies.

Following Drell, Loredo and Wasserman
\cite{DLW} we will assume that the intrinsic luminosities of SNe Ia
scale as a power of $1+z$ as a result of evolution. This model introduces a
continuous magnitude shift of the form $\beta\ln(1+z)$ to the SNe Ia sample.
Eq.~(\ref{chi2}) then becomes
\be\label{chi2-2}\chi^2(M,f)=\sum^{60}_{i=1}{(\meff{B,i}-g_i-\nu-\beta\ln
(1+z_i))^2\over\si_i^2}\ . \ee
The parameter $\beta$ will be assumed to have a Gaussian prior distribution
with mean $\beta_0$ and standard deviation $b$. Physically the parameter
$\beta_0$ represents a redshift-dependent evolution of the peak luminosity
of the supernovae sources, which might be expected to arise as a result of
the chemical evolution of the environment of the supernovae progenitors as
abundances of heavier elements increase with cosmic time. Ultimately, one
should hope to account for this evolution with astrophysical modelling of
the supernovae explosions \cite{Arn}. The parameter $b$ would then account for
a local variability in the supernovae environments between regions of
individual galaxies at the same redshift which are richer or poorer in metals,
or with progenitor populations of different ages and masses etc.

We now have two parameters to marginalize over, $\nu$ and $\beta$.
As in the case of models with no evolution, we will
marginalize over $\nu$ using a flat prior. We use a Gaussian prior for $\beta$
with mean $\beta_0$ and standard deviation $b$, so that
\be\label{prior}p(\beta')={1\over b\sqrt{2\pi}}\e^{-\beta'^2/2b^2}\ee
where
\be\beta'=\beta-\beta_0\ .\ee

The marginal likelihood is calculated by multiplying the prior (\ref{prior})
by the likelihood resulting from Eq.~(\ref{calL}), and integrating over
$\beta$. The resulting likelihood is
\bea{\cal L}(M,\ff)&=&{1\over\Delta\nu}\int{d\beta'\;\int{d\nu\;p(\beta')\e^{-
\chi^2/2}}}\nonumber\\ &=&{s\bsi\sqrt{2\pi}\over\Delta\nu b}\e^{-Q/2}
\label{calL2}\eea
where
\bea
\label{Q2}Q(M,\ff)&=&-{\hat{\beta}^2\over\bsi^2}+\sum^{60}_{i=1}{(h_i-s^2
\HH)^2\over\si_i^2}\ , \\
h_i(z_i;M,\ff)&=&\meff{B,i}-g_i-\beta_0\ln(1+z_i)\ , \\
\HH(M,\ff)&=&\sum^{60}_{i=1}{h_i\over\si_i^2}\ .
\eea
The conditional best-fit of $\beta'$ is given by
\bea
\hat{\beta}(M,\ff)&=&\bsi^2\sum^{60}_{i=1}{h_i[\ln(1+z_i)-s^2\GG]\over\si_i^2}
\ , \\
\GG&=&\sum^{60}_{i=1}{\ln(1+z_i)\over\si_i^2}\ ,
\eea
and the $\beta'$ uncertainity, $\bsi$, is given by
\be
{1\over\bsi^2}={1\over b^2}-s^2 \GG^2+\sum^{60}_{i=1}{\ln^2(1+z_i)\over
\si_i^2}\ .
\ee
Although $\bsi$ is independent of $M$ and $\ff$, the marginal likelihood is
no longer proportional to the profile likelihood because $Q$ is now given by
(\ref{Q2}) rather than by the chi-square type statistic (\ref{Q1}).

We have performed a detailed numerical analysis on the P98 dataset, varying
$\beta_0$, $b$ and $w_i$. As a result we find a best-fit value of $\beta_0
\equiv\beta^*\simeq0.414$, for $w_i=1.5$ in the PNGB models.
This would correspond to supernovae being
intrinsically dimmer by $0.17$ magnitudes at a redshift of $z=0.5$, which is
an effect of the typical order of magnitude being addressed in current
attempts to better model the supernova explosions \cite{Arn}. Furthermore,
we find that inclusion of a non-zero variance, $b^2$, does not alter the
prediction of the best-fit value of $\beta_0$, although it naturally does lead
to a broadening of the areas of parameter space included at the $2\si$ level.
There is relatively little broadening of the region of parameter values
included at the $1\si$ level in the $\beta_0=\beta^*$ plane, however, which
is no doubt a consequence of the steepness of the $\qn$ contours in
Fig.\ \ref{qn}(a) in the area corresponding to region II.

In Figs.\ \ref{SnIa-evolv1} and \ref{SnIa-evolv2} we display the joint credible
regions for $M$ and $\ff$, for two slices through the 3-dimensional $(M,\ff,
\beta_0)$ parameter space for $w_i=1.5$: (a) the best-fit case $\beta_0=\beta^
*$; and (b) $\beta_0=0$. Analogously, the best-fit case is shown in Fig.\
\ref{SnIa-evolv3} for $w_i=0.2$. We see from Fig.\ \ref{SnIa-evolv2} that once
the likelihood is normalized relative to $\beta^*$ no regions remain in the
$\beta=0$ parameter plane which are admitted at the $1\si$ level when $b=0$.
Furthermore, even when a non-zero standard deviation, $b$, is included, region
II of the $(M,\ff)$ parameter plane is favoured at the $1\si$
level, in contrast to Fig.\ \ref{SnIabeta0}.

\begin{figure}[htp] \centering \leavevmode\epsfysize=7cm
\epsfbox[80 0 400 340]{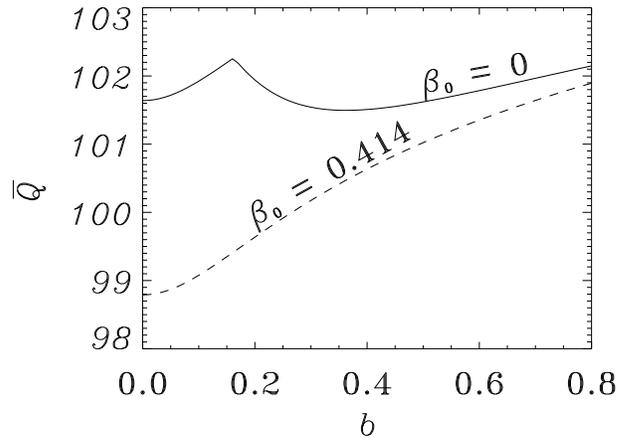}
\caption[bval]{\label{bval} Variation of the least value of $\bar Q=-2\ln
\left({\cal L}\Delta\nu\right)$ as a function of the standard deviation $b$ of
the prior distribution for $\beta_0$ for values $\beta_0=0$ and $\beta_0=
\beta^*$.}
\end{figure}

We have also undertaken an analysis of the models with $\beta_0=0$ but
variable $b$, similarly to the study of Ref.\ \cite{DLW}. In that case,
we once again find that region II of the $(M,\ff)$ parameter plane is
admitted at the $1\si$ level if $b=0.25$ or $b=0.5$. The dependence of the
value of
\be \bar Q \equiv -2\ln\left({\cal L}\Delta\nu\right)=
Q-2\ln\left(\sqrt{2\pi}\,s\bsi\over b\right)
\ee
on the value of $b$ is displayed in Fig.\ \ref{bval}, for $\beta_0=0$ as
compared with the best-fit case $\beta_0=\beta^*$. The quantity $\bar Q$ is
analagous to the chi-square statistic of the maximum likelihood method.
We see that the $\beta_0=0$ models favour a non-zero value of $b\simeq
0.36$ by a very small margin as compared to the $b=0$ case. For $b=0.36$ the
points of greatest likelihood lie mainly in region II, in contrast to the
$b=0$ case in Fig.\ \ref{SnIabeta0}.

Varying the inital condition $w_i$ does not appear to affect the best-fit
value of $\beta_0$ significantly. For $w_i=0.2$ (cf.,Fig.~\ref{SnIa-evolv3}),
for example, the best-fit value was $\beta^*=0.435$, a difference of 5\% from
the $w_i=1.5$ case. Furthermore, the numerical value of the least value of
$\bar Q$ was only 0.2\% greater in the $w_i=0.2$ case. We have not attempted
to find a best-fit value for $w_i$.

\section{Constraints from Lensing Statistics}

\begin{figure*}[htp] \centering\leavevmode\vskip2pt
\epsfysize=8cm\epsfbox{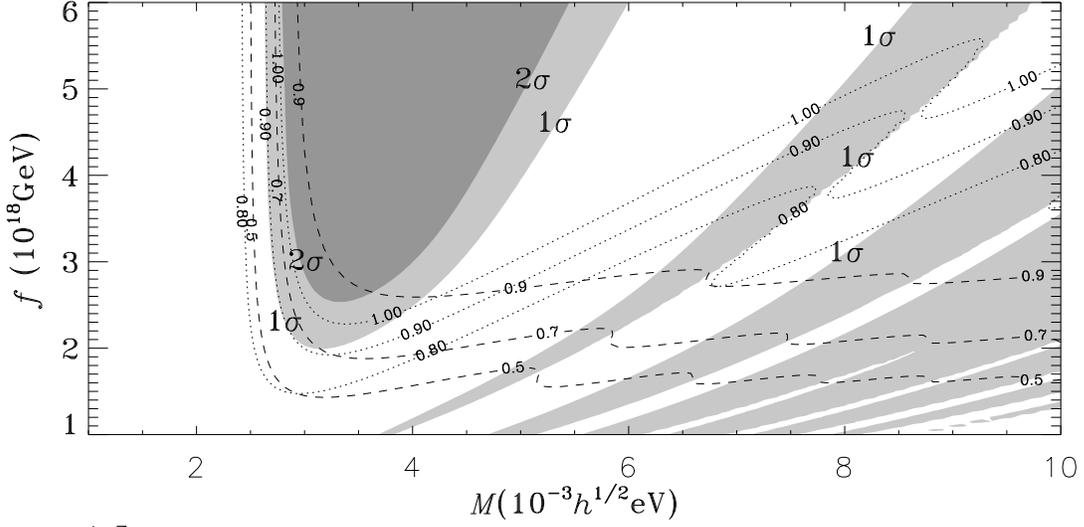}\vskip2pt
\epsfysize=8cm\epsfbox{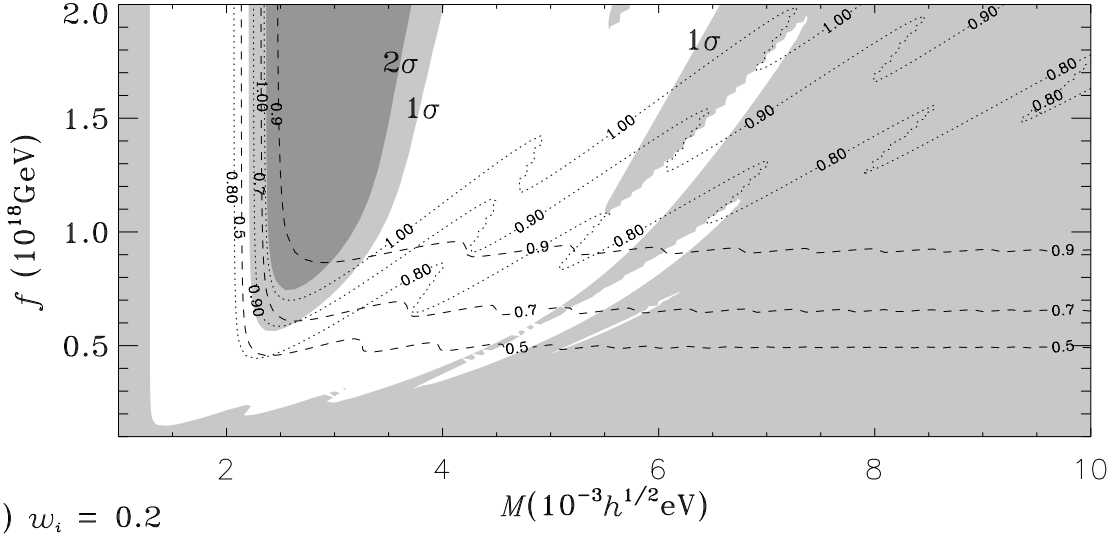}
\caption[lenstat]{\label{lenstat}
Confidence limits from gravitational lensing statistics: (a) $w_i=1.5$; (b)
$w_i=0.2$. Parameter values excluded at the 95.4\% level are darkly shaded,
while those excluded at the 68.3\% level are lightly shaded. For reference,
contours of $\Omp$ and $H_0t_0$ are superposed as dashed and dotted lines
respectively.} \end{figure*}
Gravitational lensing of distant light sources due to the accumulation of
matter along the line of sight provide another relatively sensitive constraint
on the cosmological models of interest. For cosmology the situation of most
interest is the lensing of high luminosity quasars by intervening galaxies.
The abundance of multiply imaged quasars and the observed separation
of the images to the source puts constraint on the luminosity--redshift
relation and hence the model parameters. Basically, if the volume of space
to a given redshift is larger then on average one can expect more lensing
events. This leads to a statistical test, which has been used to put bounds
on $\LA$ \cite{K95,Fal,Hel} and to test properties of some decaying $\LA$ or
quintessence models \cite{WM,lsQ}.

Gravitational lensing statistics are useful since they provide a tests which
potentially provides opposing constraints to those obtained from supernovae
magnitude--redshift tests. In particular, in the case of models with a vacuum
energy provided by a cosmological constant, the high redshift supernovae have
been interpreted as favouring relatively large values of $\Om_\LA$ - Perlmutter
\etal\ \cite{P98} give a value of $\Om_\LA=0.72$ at $1\si$ level - whereas the
gravitational lensing data leads to upper bounds on $\Om_\LA$: Kochanek
\cite{K95} quotes $\Om_\LA<0.66$ at the $2\si$ level. A combined likelihood
analysis has been performed by various authors \cite{FW2,Hel,WM}.

Gravitational lensing constraints on the PNGB models have been very recently
given by Frieman and Waga \cite{FW2} for $w_i=1.5$. However, Frieman and Waga
considered a more retricted range of parameter space, $M<0.005h\w{eV}$, since
they did not consider the possibility of source evolution in the case of the
type Ia supernovae and therefore took values of $M>0.005h\w{eV}$ to be ruled
out. We wish to extend the range of $M$ for the gravitational lensing
statistcs to consider parameter values corresponding
to region II in the supernovae constraint graphs, Figs.\ \ref{SnIabeta0},
\ref{SnIa-evolv2}, so as to compare the constraints from different tests.

We have thus simply followed the calculation described by Waga and Miceli
\cite{WM}, who performed a statistical lensing analysis of optical sources
described earlier by Kochanek \cite{K95}. They used a total of 862 ($z>1$)
high luminosity quasars plus 5 lenses from seven major optical surveys
\cite{HLQs}. (Another alternative not considered here is to analyse data from
radio surveys -- see, e.g., \cite{Fal,Hel}.) Undertaking a similar analysis
for the increased parameter range, we arrive at Fig.~\ref{lenstat}, which
shows the 68.3\% and 95.4\% joint credible regions for $M$ and $\ff$, for two
values of $w_i$. We refer the reader to Refs.\ \cite{K95,WM} for details of
the calculation. The only regions of parameter space excluded at the $2\si$
level turn out to be areas of parameter space for which the deceleration
is presently negative (cf., Fig.\ \ref{qn}), with the scalar field still
commencing its first oscillation at the present epoch.

\section{Discussion}

Let us now consider the overall implications of the constraints observed
above.

Firstly, since empirical models with source evolution do appear to fit the
data somewhat better, it would appear that we do have weak evidence for an
underlying evolution of the peak luminosity of the type Ia supernovae
sources, at least in the context of the PNGB quintessence models. It might
be interesting to compare the case of other quintessence models, or
the case of a cosmological constant. However, the PNGB model is qualitatively
different from such models since its final state corresponds to one in
which the ultimate
destiny of the universe is to expand at the same rate as a spatially flat
Friedmann--Robertson--Walker model, rather than to undergo an accelerated
expansion. This is of course precisely why we chose the PNGB models as the
basis of our investigation, rather than models in which a late--time
accelerated expansion had been built in by hand. If we wish to test the
hypothesis that the faintness of the type Ia supernovae is at least
partly due to an intrinsic variation of their peak luminosities -- which
is a very real possibility in view of the results of \cite{Rie1} -- then
a quintessence model which possesses a variety of possibilities for the
present--day variation of the scale factor is probably the best type of
model to investigate.

If only supernovae luminosity distances, (cf., Fig.\ \ref{SnIa-evolv1},
and gravitational lensing statistics, (cf., Fig.\ \ref{lenstat}), are compared
then we see that there is a remarkable concordance between the two tests --
region II of Fig.\ \ref{SnIa-evolv1} coincides with a region included at even
the $1\si$ level in Fig.\ \ref{lenstat}. This is perhaps not surprising, since
in view of Fig.\ \ref{qn} region II corresponds to parameter values for which
the present day universe has already undergone almost one complete oscillation
of the scalar field about the final critical point $\Co_2$ of Fig.\
\ref{rhgm0}. It is thus already well on the way towards its asymptotic
behaviour, which closely resembles that of a standard spatially flat
Friedmann-Robertson-Walker model.

Due to the oscillatory behaviour, parameter values in region II correspond
to models in which there has been a recent cosmological acceleration, (e.g.,
at $z\goesas0.2$), but with a $q(z)$ which changes sign three times over the
larger range of redshifts, $0<z<4$, in the quasar lensing sample, and therefore
differing significantly from Friedmann--Lema\^{\i}tre models over this larger
redshift range. Extending the SNe Ia sample to include objects at redshifts
$0.1<z<0.4$ and $z>0.85$ in substantial numbers would greatly improve the
ability to decide between models in regions I and II.

There is some cause for concern, however, if we consider the favoured values
of $\Omp$ and $H_0t_0$. In the case of the models with empirical evolution of
the supernovae sources, we always found that the best-fit parameter values
occured at the $\Omp\rarr1$ boundary of the ($M,\ff$) parameter space. The
overwhelming evidence of many astronomical observations over the past two
decades \cite{OS} would tend to indicate that $\Om_{m0}\simeq0.2\pm0.1$,
indicating that a vacuum energy fraction of $\Omp\goesas0.7$--$0.8$ is
desirable, and $\Omp\lsim0.9$ in any case. Although parameter values with
$\Omp<0.8$ certainly fall within both the 2$\si$ and 1$\si$ portions of region
II of Fig.\ \ref{SnIa-evolv1}, for all values of $b$, there are potentially
serious problems if we wish to simultaneously obtain large values of $H_0t_0$.
In view of recent estimates of the ages of globular clusters \cite{Krauss},
a lower bound of 12 Gyr for the age of the Universe appears to currently
indicated. With $h\simeq0.65$ this would require $H_0t_0\gsim0.8$. For
$w_i=1.5$, parameter values with $H_0t_0>0.8$ coincide with values $\Omp\gsim
0.9$ in region II, which is phenomenologically problematic.

The tension between the values of $\Omp$ and $H_0t_0$ is somewhat mitigated
for lower values of $w_i$. For $w_i=0.2$, for example, we see from Figs.\
\ref{SnIa-evolv3} and \ref{lenstat}(b) that the $\Omp=0.7$ and $H_0t_0$ meet
in Region II, and there is a small region of parameters there with $0.7\lsim
\Omp\lsim0.9$ and $H_0t_0\gsim0.8$, which is also consistent with the other
cosmological tests.

Even if the supernovae sources undergo evolution it is clear that
parameter values in region I of \ref{SnIabeta0} which are favoured in the
absence of evolution of peak SnIa luminosities, are still included at the
$2\si$ level in the models with evolution, in view of Fig.\ \ref{SnIa-evolv1}.

Perhaps the most significant aspect of our results is the fact that the mere
introduction of an additional dispersion, $b>0.17$, in the peak luminosities
while leaving their mean value fixed (cf., Fig.\ \ref{bval}), gives rise to a
change in the best-fit region of parameter space from region I to region II.
(See \cite{dlw} for further details.)
One would imagine that an increased dispersion is
likely to be a feature of many models of source evolution, even if evolutionary
effects are of secondary importance. Thus even if the empirical models with
non-zero $\beta$ are somewhat artificial, more sophisticated scenarios could
well lead to similar changes in regard to the fitting of cosmological
parameters in the PNGB model.

Much tighter bounds on the parameter space of quintessence models, including
the present model, will be obtained over the next decade as more supernovae
data is collected. What we wish to emphasize, however, is that an effective
vacuum energy which is cosmologically significant at the present epoch should
not simply be thought of in terms of a ``cosmic acceleration''. A dynamical
vacuum energy with a varying effective equation of state allows for many
possibilities for the evolution of the universe, and overly restrictive
assumptions, such as equating quintessence to models with a late period of
continuous cosmological acceleration, should be avoided. If detailed
astrophysical modeling of type Ia supernovae explosions ultimately shows that
the dimness of distant supernova events is largely due to evolutionary effects,
it does not spell the end for cosmologies with dynamical scalar fields.

\section*{Acknowledgments}

We would like to thank Chris Kochanek for supplying us with the gravitational
lensing data, Elisa di Pietro, Don Page and Ioav Waga for helpful discussions
about various aspects of the paper, and the Australian Research Council for
financial support.

\def\PRL#1{Phys.\ Rev.\ Lett.\ {\bf#1}} \def\PR#1{Phys.\ Rev.\ {\bf#1}}
\def\ApJ#1{Astrophys.\ J.\ {\bf#1}} \def\AsJ#1{Astron.\ J.\ {\bf#1}}
\def\CQG#1{Class.\ Quantum Grav.\ {\bf#1}} \def\Nat#1{Nature {\bf#1}}
\def\JMP#1{J.\ Math.\ Phys.\ {\bf#1}} \def\NP#1{Nucl.\ Phys.\ {\bf#1}}
\def\PL#1{Phys.\ Lett.\ {\bf#1}} \def\AsAp#1{Astron.\ Astrophys.\ {\bf#1}}
\def\MNRAA#1{Mon.\ Not.\ Roy.\ Astron.\ Soc.\ {\bf#1}}


\begin{references}
\bibitem{DLWnote}
Our quantity $\nu$ is identical to the quantity $\gm$ given in Appendix B
of \cite{DLW}.
\end{references}
\end{document}